\RequirePackage[T1]{fontenc}
\documentclass[10pt,conference,letterpaper]{IEEEtran}
\usepackage{amsmath,amssymb,amsthm}
\usepackage{booktabs,array,multirow}
\usepackage{graphicx}
\usepackage[table]{xcolor}
\usepackage{cite}
\usepackage{url}
\usepackage{microtype}
\usepackage[hidelinks]{hyperref}
\usepackage{algorithm}
\usepackage{algpseudocode}
\newcommand{\quiver}{QuIVer}

\newcommand{\pc}{\operatorname{popcnt}}

\newcommand{\R}{\mathbb{R}}
\newcommand{\missing}{\textemdash}
\makeatletter

\newsavebox{\quiverTableBox}
\def\quiverLayoutIssues{}
\newenvironment{quivercheckedtable}[1]{%
  \def\quiverTableName{#1}%
  \begin{lrbox}{\quiverTableBox}%
}{%
  \end{lrbox}%
  \ifdim\wd\quiverTableBox>\linewidth
    \xdef\quiverLayoutIssues{\quiverLayoutIssues\space 表格\quiverTableName=\the\wd\quiverTableBox\space（上限\the\linewidth）；}%
  \fi
  \usebox{\quiverTableBox}%
}
\AtEndDocument{%
  \@ifundefined{b@lewis2020rag}{}{%
    \ifx\quiverLayoutIssues\@empty\else
      \PackageError{QuIVerFormat}{IEEE排版检查未通过：\quiverLayoutIssues}{请重新安排内容，保留IEEE正文字号、行距和页边距。}%
    \fi
  }%
}
\makeatother

\graphicspath{{figures/}}
\newcommand{\theoryfigure}[1]{\includegraphics[width=\textwidth]{#1.png}}
\title{QuIVer: Efficient Graph Indexing with Two-Bit Construction and Navigation}
\author{
\IEEEauthorblockN{Wenxuan Xiao\IEEEauthorrefmark{1}, Peidong Zhu\IEEEauthorrefmark{1}, Chengcheng Li\IEEEauthorrefmark{1}, Xu Cao\IEEEauthorrefmark{2}}
\IEEEauthorblockA{\IEEEauthorrefmark{1}Changsha University, Changsha, China\\
daflyflowers@gmail.com, zpd@ccsu.edu.cn, qq1330494624@outlook.com}
\IEEEauthorblockA{\IEEEauthorrefmark{2}Astrmira Tech.\quad x.cao@astrmira.com}}
\begin{document}
\maketitle
\begin{abstract}
High-dimensional vector indices must balance construction time, memory, and query throughput. Two-bit comparisons reduce computation and storage, but preserving useful graph connections and search candidates requires scores suited to each stage. We present QuIVer, a graph index that uses a training-free sign--magnitude representation for construction and navigation while reserving original vectors for bounded final reranking. Construction discovers candidates with Hamming scores, orders them with an integer proxy, and defers reverse-edge pruning across parallel insertions. Search uses reconstruction norms and bitwise inner products to order both the frontier and retained candidates, without query calibration or an exact anchor. Experiments cover 21 datasets and compare 30 external baseline modes on three core workloads. QuIVer's validation-selected 95\% configurations lie on the measured Pareto frontiers of achieved recall, throughput, index preparation time, and memory, whether memory counts navigation state alone or also the original vectors. On Cohere-5M, QuIVer reaches 95.90\% Recall@10 at 3,075 single-thread queries/s, $11.74\times$ the throughput of Lucene BBQ with FP32 reranking at 95.83\% recall. Against SymphonyQG at 95.87\%, it retains 81\% of single-thread and 97\% of 32-thread throughput while reducing index preparation time by 54\%, total logical memory by 47\%, and navigation storage by 87\%. An SSD mode with INT8 reranking serves Cohere-5M at about 5 GiB of steady-state process RSS. Controlled crossovers and candidate-coverage diagnostics connect these results to construction and retention quality. Across 21 datasets, high recall holds for most text and image embeddings, while non-negative descriptors, cross-modal queries, and random data expose the limits of the representation.
\end{abstract}

\section{Introduction}
\label{sec:intro}
Learned embeddings support retrieval-augmented generation~\cite{lewis2020rag}, recommendation and search~\cite{sobot2023voyager}, code assistance~\cite{fu2025copilot}, multimodal retrieval~\cite{google2026embedding}, and protein-structure search over 214 million predicted structures~\cite{segura2026structure}. Across these applications, embeddings become reusable data objects stored and accessed through vector-similarity queries.

Serving these collections couples three system costs: construction time, memory, and query throughput. A million 1,536-dimensional FP32 vectors occupy approximately 5.72 GiB before graph edges, identifiers, or query workspaces are stored. A graph search repeatedly fetches small, irregular subsets of this representation, while graph construction performs many more comparisons to discover and prune candidate edges. Reducing the representation involved in those comparisons can therefore improve both index preparation and recurring search cost.

Compressed graph indices already exploit this opportunity. Locally adaptive vector quantization (LVQ) supports graph construction and search with compact scalar codes~\cite{aguerrebere2023lvq}. NGT-QG and SymphonyQG arrange quantized neighborhoods for batched SIMD scoring~\cite{ngt,gou2025symphony}, and Lucene's Better Binary Quantization (BBQ) combines very small database codes with higher-precision query codes~\cite{trent2024bbq,veasey2024osq}. The relevant design space thus includes the graph, its scoring function, and the memory layout.

Aggressive compression introduces a particular tension within that design space. A cheaper score is useful only if it saves more work than its errors create. Approximate scores determine which vertices are expanded, which candidates survive a bounded beam, and which survivors receive expensive exact evaluation. These decisions determine the neighbors available to final reranking. Thus, the cost of low-bit search depends on both \emph{the price of each comparison} and \emph{the number and quality of the comparisons induced by the representation}.

Our insight is that construction and navigation can share two coordinate bits while using different scoring rules. Construction discovers candidates with two-plane Hamming scores and orders the candidate pool with an integer proxy. Query navigation combines these bits with per-vector reconstruction amplitudes and norms to order the frontier and candidate pool. Original vectors are accessed only for bounded final reranking. This division of precision reduces repeated construction work and navigation storage while preserving the candidates needed for accurate retrieval on semantic embeddings.

The distinction between visiting and retaining candidates is central. On a controlled Cohere-200K experiment, an unnormalized reconstruction inner product visits a set covering 96.31\% of the true top ten, but retains only 76.93\% in its final beam at width 64. Reconstruction L2 reaches 96.48\% Recall@10. This is a failure of approximate selection inside the search, even though the graph has exposed many correct neighbors. It motivates using the same norm-aware score for both expansion and candidate retention.

\begin{figure*}[!t]
\centering
\includegraphics[width=\textwidth]{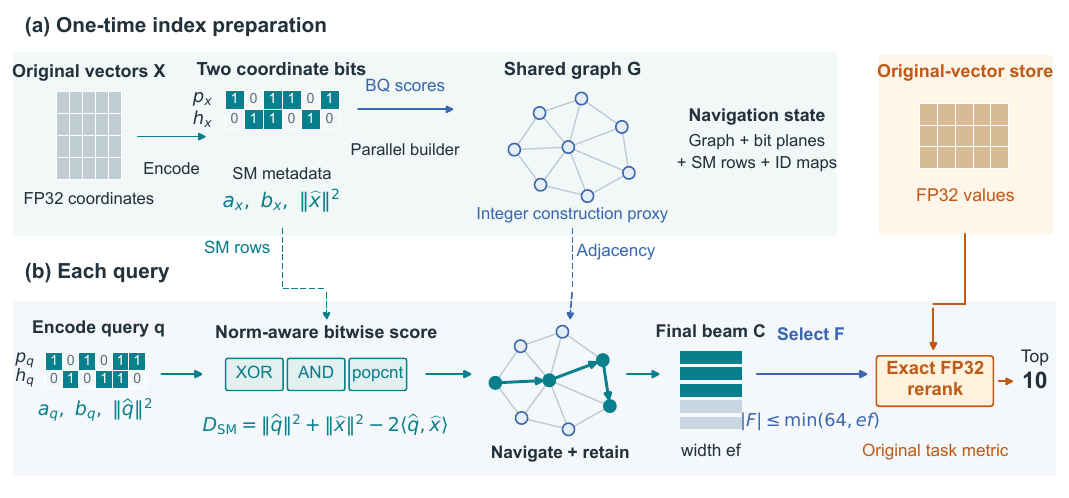}
\caption{Precision across the QuIVer lifecycle. (a) Two coordinate bits support integer BQ construction; per-vector metadata prepares SM-L2 navigation. (b) Bitwise scoring drives traversal and retention; at most $\min(64,ef)$ candidates receive FP32 reranking. Dashed links supply persistent navigation state, and amber supplies original values. Bit patterns and graph layouts are schematic.}
\label{fig:pipeline}
\end{figure*}

We implement this design as \quiver{} in TriviumDB. Parallel insertion and deferred reverse-edge pruning accelerate construction; packed bit planes, population-count kernels, shared adjacency, and reusable thread-local state accelerate queries (Figure~\ref{fig:pipeline}). The resulting index occupies a measured Pareto-frontier position on all three core workloads when achieved recall, throughput, preparation time, and memory are considered jointly, whether or not memory includes the original vectors. On Cohere-5M, it retains 97\% of SymphonyQG's 32-thread throughput with 54\% less preparation time and 47\% less total logical memory. An SSD mode keeps this navigation path in memory and reads INT8-quantized vectors for bounded reranking, serving Cohere-5M at about 5 GiB of steady-state process RSS.

We make three contributions:
\begin{itemize}
    \item A two-bit graph-index design that assigns inexpensive integer scoring to construction, reconstruction-aware scoring to navigation and retention, and bounded original-vector access to final ranking.
    \item An implementation that makes this precision split effective through parallel construction, deferred pruning, bitwise scoring, shared adjacency storage, and reusable query state, with separately measured preparation and memory costs, and an SSD mode that fetches INT8-quantized reranking rows in one bounded batch per query.
    \item Evidence of Pareto-efficient indexing on three core workloads, supported by comparisons with 30 external modes, a 21-dataset applicability study, held-out queries, and controlled experiments separating construction, navigation, candidate retention, and reranking.
\end{itemize}

\section{Two-Bit Graph Indexing}
\label{sec:design}

\subsection{Problem and Precision Budget}
Let $X=\{x_1,\ldots,x_n\}\subset\R^d$ be the database and $q\in\R^d$ a query. We return $k$ approximate nearest neighbors under the task metric. The primary task is cosine search: database and query vectors are normalized, and exact ranking uses their original FP32 values. We also evaluate native squared Euclidean distance on SIFT and GIST, retaining their original magnitudes. A graph $G$ stores bounded outgoing neighbor lists. Search maintains a candidate beam of capacity $e$, denoted by \texttt{ef} in the implementation, and reranks at most $B$ candidates.

The main configuration uses $k=10$ and
\begin{equation}
    B_{\mathrm{eff}}=\min(B,e),\qquad B=64,\quad e\ge k.
    \label{eq:budget}
\end{equation}
Increasing $e$ spends more navigation work; increasing $B$ spends more original-vector work. The two parameters control different failure modes and are evaluated separately in Section~\ref{sec:ablation}.

\subsection{Sign--Magnitude Representation}
For a vector $x$, define its mean absolute coordinate value and two binary planes:
\begin{equation}
\begin{split}
    \tau_x&=\frac{1}{d}\sum_{j=1}^d |x_j|,\\
    p_{xj}&=\mathbf{1}[x_j\ge0],\qquad
    h_{xj}=\mathbf{1}[|x_j|\ge\tau_x].
\end{split}
\label{eq:bits}
\end{equation}
Write $\sigma_{xj}=2p_{xj}-1$. The sign plane distinguishes positive and negative coordinates; the magnitude plane distinguishes weak and strong coordinates relative to the vector's own scale. Each plane is packed into machine words. The two planes require $2d$ bits before word-alignment padding.

To recover scale information for query scoring, compute the mean absolute value within each group:
\begin{equation}
\begin{split}
 a_x&=\operatorname{mean}_{h_{xj}=0}|x_j|,
 \quad b_x=\operatorname{mean}_{h_{xj}=1}|x_j|,\\
 \delta_x&=b_x-a_x.
\end{split}
\end{equation}
An empty group contributes zero. The reconstructed coordinate is
\begin{equation}
    \widehat{x}_j=\sigma_{xj}(a_x+\delta_x h_{xj}).
    \label{eq:reconstruction}
\end{equation}
The scalar metadata are derived independently for each vector. Figure~\ref{fig:theory-geometry}(a) illustrates the resulting four-level reconstruction and its coordinate residuals. No codebook, rotation, training queries, or ground-truth neighbors are needed. The query is encoded by the same rule. For a zero vector, both reconstruction amplitudes are zero, so its native-L2 reconstruction remains well defined.

\subsection{Low-Cost Construction}
\label{sec:build}
Construction uses a fixed-amplitude version of the code,
$u_{xj}=\sigma_{xj}(1+h_{xj})\in\{-2,-1,1,2\}$. Its integer score is
\begin{equation}
    D_{\mathrm{BQ}}(x,y)=4d-\sum_j u_{xj}u_{yj}.
    \label{eq:bq}
\end{equation}
Positive and negative contributions have weights 1, 2, or 4, according to the two magnitude bits. Let $M=p_x\oplus p_y$, $G_1=\neg(h_x\lor h_y)$, $G_2=h_x\oplus h_y$, and $G_4=h_x\land h_y$, with complements restricted to valid coordinates. Six population-count reductions compute the signed sum as $\sum_{w\in\{1,2,4\}}w[\pc(G_w\land\neg M)-\pc(G_w\land M)]$. Equation~\eqref{eq:bq} includes the offset and same-sign contributions. Its self-score depends on the magnitude pattern, so we use it as an ordering and pruning proxy.

All database codes and graph slots are allocated before insertion. The builder processes batches of 256 vertices with a fixed-size worker pool. A construction beam starts from vertex zero and uses the inexpensive two-plane Hamming score
\begin{equation}
 H(x,y)=\pc(p_x\oplus p_y)+\pc(h_x\oplus h_y)
\end{equation}
to discover candidates. The candidate set is augmented by deterministic pseudorandom samples from all $n$ pre-encoded vectors, including vertices awaiting insertion, deduplicated, and sorted with $D_{\mathrm{BQ}}$. The sample count is
$\max(\min(e_c,128),2R)$, where $e_c$ is the construction beam width and $R$ is the final degree bound. The evaluated setting has $e_c=128$ and $R=2m=64$, where $m=32$ is the builder's base connectivity parameter.

Neighbor selection combines a pairwise rejection test with ordered backfill. A candidate $y$ for source $x$ is rejected during the first pass if an already selected neighbor $z$ satisfies
\begin{equation}
    D_{\mathrm{BQ}}(y,z)<\alpha D_{\mathrm{BQ}}(x,y),
    \qquad \alpha=1.2.
    \label{eq:prune}
\end{equation}
Remaining slots are filled from the unselected candidates in $D_{\mathrm{BQ}}$ order. With $S(x,y)=\sum_j u_{xj}u_{yj}$, the test rejects $y$ whenever $S(y,z)>1.2S(x,y)-0.8d$, so the offset makes it reject almost every candidate after the nearest one. In a traced production build of Cohere-1M, 98.44\% of the final edges come from backfill, and each vertex keeps one or two diversity-selected edges. Each list therefore holds essentially the $R$ candidates nearest under $D_{\mathrm{BQ}}$ in its pool, which includes reverse edges. Section~\ref{sec:mechanism} measures rules that admit more diverse edges.

The builder adds reverse edges under striped locks. Repruning on every reverse insertion repeatedly scores almost the same neighbor list. We instead reserve capacity for $2R$ neighbors, append new reverse edges while capacity remains, and prune when a list fills. A final parallel pass removes self-loops and duplicates from every oversized list, sorts it by $D_{\mathrm{BQ}}$, and applies the same selection and backfill to keep $R$ neighbors. The query representation shares this adjacency allocation, including its reserved capacity. Section~\ref{sec:system} measures the effect of deferred pruning against an eager variant.

Parallel scheduling affects which intermediate edges are visible to construction beams. Each query comparison reuses its constructed graph; repeated builds are reported separately.

\subsection{Norm-Aware Navigation with Bitwise Scores}
\label{sec:score}
The query path evaluates reconstructed squared distance,
\begin{equation}
    D_{\mathrm{SM}}(q,x)=
    \|\widehat q\|^2+\|\widehat x\|^2
    -2\langle\widehat q,\widehat x\rangle .
    \label{eq:l2}
\end{equation}
For cosine tasks, the original vectors are normalized before encoding, but their reconstructions generally do \emph{not} have identical norms. Keeping the norm term prevents changes in reconstruction magnitude from acting as an unintended ranking preference.

The inner product in Equation~\eqref{eq:l2} does not require decoding all coordinates. Let $M=p_q\oplus p_x$, and define
\begin{align}
 c_0&=d-2\pc(M),\\
 c_x&=\pc(h_x)-2\pc(M\land h_x),\\
 c_q&=\pc(h_q)-2\pc(M\land h_q),\\
 c_{qx}&=\pc(h_q\land h_x)-2\pc(M\land h_q\land h_x).
\end{align}
Then
\begin{equation}
\begin{split}
\langle\widehat q,\widehat x\rangle
={}&a_q a_x c_0+a_q\delta_x c_x\\
&+\delta_q a_x c_q+\delta_q\delta_x c_{qx}.
\end{split}
\label{eq:bitdot}
\end{equation}
The counts of strong coordinates are stored or computed once. Five masked population-count reductions provide the remaining terms. The squared reconstruction norm is
\begin{equation}
    \|\widehat x\|^2=(d-s_x)a_x^2+s_xb_x^2,
    \quad s_x=\pc(h_x),
\end{equation}
and is cached in the SM row. The score therefore combines bitwise reductions with a constant amount of scalar floating-point arithmetic.

\begin{algorithm}[t]
\caption{QuIVer query with bounded exact reranking}
\label{alg:query}
\begin{algorithmic}[1]
\Require Query $q$, shared graph $G$, entry $v_0$, width $e$, budget $B$
\State Encode $q$ and compute its reconstruction norm
\State Reset the reusable frontier, beam, and visited state
\State Score and insert $v_0$ into the frontier and beam
\While{an eligible unexpanded frontier candidate remains}
    \State Extract the best frontier vertex $v$
    \If{the beam is full and $v$ is worse than its worst entry}
        \State \textbf{break}
    \EndIf
    \For{each unvisited $x$ in $G[v]$}
        \State Mark $x$ visited and evaluate $D_{\mathrm{SM}}(q,x)$
        \If{$x$ improves the beam or the beam has space}
            \State Insert $x$ into the frontier and beam
            \State Retain the best $e$ beam entries
        \EndIf
    \EndFor
\EndWhile
\State Select the best $\min(B,e)$ surviving candidates by $D_{\mathrm{SM}}$
\State Fetch their original vectors and evaluate the exact task metric
\State \Return the exact top $k$ among the selected candidates
\end{algorithmic}
\end{algorithm}

Algorithm~\ref{alg:query} uses $D_{\mathrm{SM}}$ for both frontier order and candidate retention. All evaluated production graphs use the fixed entry $v_0=0$. Beam evictions do not truncate the frontier; the loop stops once the best frontier vertex is worse than the worst entry of a full beam. The final SM ordering selects the candidates that consume the FP32 budget. A callback maps their graph identifiers to slots in the original-vector store during final reranking.

\subsection{Layout and Execution}
Packed rows contain the two bit planes, reconstruction amplitudes, strong-bit count, and cached norm. For $d=768$, a 256-byte row holds two 96-byte planes, four FP32 scalars (two amplitudes, the norm, and a correction coefficient unused by SM-L2), and the count, padded to a 64-byte boundary. A dimension-aware path handles the other evaluated dimensions, including dimensions that are not multiples of 64. Tail masks exclude padded coordinates. Runtime dispatch selects available vector population-count kernels, with scalar fallback.

Scoring neighboring candidates in small interleaved groups exposes independent memory operations. Prefetching brings upcoming rows into cache. Per-thread workspaces reuse the frontier, candidate storage, and visited bitset instead of allocating them for every query. The graph is shared by construction and SM search; the database's original BQ codes remain allocated alongside the SM representation. We include both code stores in the memory measurements.

SM metadata installation is a separate pass over the input. It is required before the evaluated query path is ready, and the measured reload path reinstalls this state. We report this cost next to graph construction and distinguish both from database import, serialization, and loading. The evaluation uses the October 17--18 implementation snapshot with native-L2 and low-width query support.

\subsection{SSD-Resident Reranking}
\label{sec:ssd}
Navigation reads only the graph and SM rows, so the original vectors can leave memory. An SSD mode, implemented in a TriviumDB research branch, stores each vector as a 1 KiB row of INT8 codes with min--max scales per 128-coordinate group and a norm correction. Once navigation fixes its at most $B$ candidates, the query submits all of their reads together through four independent io\_uring shards with \texttt{O\_DIRECT}, and scores the rows with integer dot products against a query quantized once with a global scale. Per-shard FIFO caches totaling 128 MiB and a 32 MiB in-flight buffer bound the memory used for rows. Disk-resident graph indices read neighborhoods from SSD during traversal~\cite{jayaram2019diskann,wang2024starling,guo2025pipeann}; this mode keeps traversal in memory and issues one batch of at most $B$ independent reads per query.

\section{Decision Quality in Two-Bit Search}
\label{sec:analysis}

\begin{figure*}[!t]
\centering
\theoryfigure{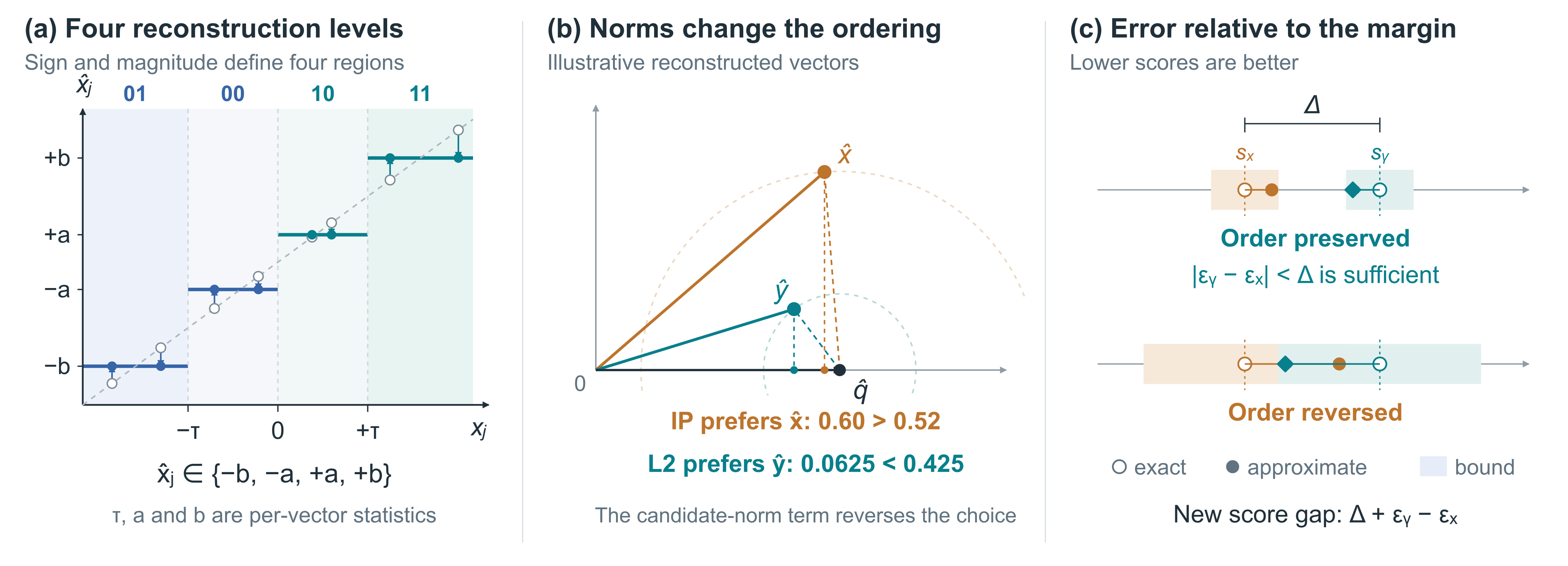}
\caption{Low-bit decision geometry. (a) Four reconstruction levels and coordinate residuals. (b) With $\widehat q=(0.8,0)$, $\widehat x=(0.75,0.65)$, and $\widehat y=(0.65,0.2)$, inner product favors $x$ but squared L2 favors $y$. (c) For lower-is-better scores, the exact gap $\Delta=s_y-s_x>0$ becomes $\Delta+\epsilon_y-\epsilon_x$ under errors $\epsilon_i=\widehat s_i-s_i$. Overlapping error bounds permit a reversal. Coordinates and bands are illustrative.}
\label{fig:theory-geometry}
\end{figure*}

\subsection{Local Scores and Global Search Work}
For a query, write $N_s$ for the number of approximate scores, $N_e$ for the number of expanded vertices, and $N_f$ for the number of exact reranking evaluations. A useful cost decomposition is
\begin{equation}
\begin{split}
 T(q)={}&T_{\mathrm{encode}}+N_s c_{\mathrm{SM}}\\
       &+T_{\mathrm{graph,pool}}+N_f c_{\mathrm{FP32}},
 \quad N_f\le B_{\mathrm{eff}}.
\end{split}
 \label{eq:cost}
\end{equation}
The quantities $c_{\mathrm{SM}}$ and $c_{\mathrm{FP32}}$ include the cost of accessing their operands. $T_{\mathrm{graph,pool}}$ includes adjacency traversal, visited-state updates, and candidate maintenance. Vector population-count kernels reduce the per-block arithmetic in $c_{\mathrm{SM}}$.

Compression can change $N_s$, $N_e$, and candidate survival simultaneously. A higher-fidelity score can reduce the number of comparisons needed to reach the target recall, offsetting its greater cost per comparison.

\subsection{Why the Reconstruction Norm Matters}
Consider two candidates $x$ and $y$. Reconstructed inner product prefers $x$ when
$\langle\widehat q,\widehat x\rangle>
\langle\widehat q,\widehat y\rangle$.
Reconstructed L2 instead compares
\begin{equation}
\begin{split}
 D_{\mathrm{SM}}(q,x)-D_{\mathrm{SM}}(q,y)
 ={}&\|\widehat x\|^2-\|\widehat y\|^2\\
 &-2\langle\widehat q,\widehat x-\widehat y\rangle.
\end{split}
\label{eq:normgap}
\end{equation}
The candidate-norm difference can reverse the decision even when the original vectors all have unit norm. Figure~\ref{fig:theory-geometry}(b) isolates this effect at the reconstruction level: the candidate with the greater query projection is farther away in squared distance. This difference motivates the reconstruction-L2 and reconstruction-cosine controls in Section~\ref{sec:ablation}.

For squared Euclidean scoring, let $r_x=x-\widehat x$ and $r_q=q-\widehat q$. The score error has the exact identity
\begin{equation}
\begin{split}
 \|q-x\|^2-\|\widehat q-\widehat x\|^2
 ={}&2\langle\widehat q-\widehat x,r_q-r_x\rangle\\
 &+\|r_q-r_x\|^2.
\end{split}
\label{eq:residual}
\end{equation}
Thus, the direction of the residual relative to a comparison matters in addition to its magnitude. For a fixed pair of candidates with exact score gap $\Delta>0$, their order is preserved whenever the difference between their score errors has magnitude below $\Delta$. Figure~\ref{fig:theory-geometry}(c) contrasts preserved ordering with a possible reversal as error bounds overlap. During traversal, these comparisons govern frontier order and candidate retention. Distributional residual bounds use joint concentration assumptions, as in Bernstein-type inequalities~\cite{bernstein1924,boucheron2013concentration} and high-dimensional probability~\cite{vershynin2018highdim}. Decision-level distributional bounds and coupling results are developed by Xiao and Cao~\cite{xiao2026decisions}; this paper contributes the index design, its bitwise implementation, and the stage-wise diagnostics below.

\subsection{An Exact Decomposition of Candidate Loss}
Let $T_k(q)$ be the true top-$k$ set, $V(q)$ the set of vertices scored during traversal, $C(q)\subseteq V(q)$ the final beam, and $F(q)\subseteq C(q)$ the candidates sent to exact reranking. Assume a consistent tie convention, $|F(q)|\ge k$, and exact final scoring. Each true neighbor in $F(q)$ must appear in the returned top $k$, because fewer than $k$ database vectors can outrank it. Therefore
\begin{equation}
\begin{split}
1-\operatorname{Recall@}k(q)
={}&\frac{|T_k\setminus V|}{k}
 +\frac{|T_k\cap(V\setminus C)|}{k}\\
 &+\frac{|T_k\cap(C\setminus F)|}{k}.
\end{split}
\label{eq:decomposition}
\end{equation}
The three terms identify traversal loss, beam-retention loss, and reranking-selection loss: a true neighbor may be scored yet excluded from the final beam, or survive in the beam yet miss the exact budget. We compute these oracle coverages after each run from its visited set, retained candidates, and ground truth.

The decomposition attributes each missed true neighbor to the stage that discards it. The construction--query crossovers in Section~\ref{sec:mechanism} further separate graph quality from query scoring.

\subsection{Memory and Deployment Cost}
The logical state has three principal parts:
\begin{equation}
 M_{\mathrm{nav}}+M_{\mathrm{rerank}}+M_{\mathrm{runtime}}.
\end{equation}
$M_{\mathrm{nav}}$ includes graph capacity, original BQ codes, derived SM rows, scalar factors, and identifier mappings. $M_{\mathrm{rerank}}$ is the original-vector store: $4nd$ bytes for dense FP32 vectors. We call $M_{\mathrm{total}}=M_{\mathrm{nav}}+M_{\mathrm{rerank}}$ the total logical memory. $M_{\mathrm{runtime}}$ includes thread workspaces, object and container overhead, allocator state, and other process data; process RSS also reflects page-cache state.

For Cohere-1M, the measured persistent navigation arrays contain 192 MB of original BQ codes, 264 MB of SM-owned rows and identifiers, 516 MB of shared adjacency capacity, and an additional 16 MB of ID/slot mappings, totaling 988 MB or 0.920 GiB (MB denotes $10^6$ bytes). The BQ codes serve incremental insertion and the fallback route, and the reserved slots serve deferred pruning; a static layout without them would hold 0.503 GiB by layout arithmetic, but all measurements use the implemented layout. Each query thread additionally uses an $n$-bit visited set, or 125,000 bytes for this dataset. The 2.861 GiB original-vector store brings $M_{\mathrm{total}}$ to 3.781 GiB.

\section{Evaluation}
\label{sec:evaluation}

\subsection{Experimental Setup}
\paragraph{Hardware and execution}
All experiments run on one instance configuration: an Intel Xeon Platinum 8352V with a 32-CPU allocation, a 60 GiB memory limit, and a 56 GiB process-family guard. Query throughput is measured at 1, 8, 16, and 32 threads with fixed CPU affinity. Every index builder is configured with eight threads; Lucene adds documents serially before its eight-thread merge, and a separate experiment varies QuIVer's construction threads. Builds, timed queries, compilation, and large transfers are serialized to avoid interference. Supplementary experiments build their own indices from the same inputs and frozen configurations; their throughput and build times are compared only within each experiment.

QuIVer is compiled in release mode with native CPU targeting, link-time optimization, and one code-generation unit; the C++ wrappers we compile for SymphonyQG, HNSWlib, and VSAG use \texttt{-O3 -march=native}, and RaBitQ keeps its official SIMD dispatch. The query benchmark calls the integrated QuIVer kernel with the database's vector-pool callback. It measures ANN processing and reranking, excluding network transport, query-language parsing, and payload serialization. External methods use their native batch interfaces, with interface and timing boundaries recorded per measurement. SVS processes each query set in one call and exposes no per-request latency, so its tail latencies are unreported.

\paragraph{Datasets and splits}
Table~\ref{tab:datasets} lists all 21 datasets, including two five-million-vector collections. Three workloads have external comparisons: Cohere-1M-768, DBpedia-990K-1536, and Cohere-5M-768. The remaining datasets evaluate QuIVer's dimensionality, scale, and representation dependence. Cosine datasets are normalized, while SIFT and GIST retain native L2 distance, original magnitudes, and zero vectors. Ground truth refers to the actual evaluated database subset. In particular, truncated Arxiv, CodeSearch, and GooAQ collections use recomputed FP64 ground truth over their million-vector subsets.

Each dataset reserves 256 queries for parameter selection. The remaining 744 or 9,744 queries, depending on the dataset, form the evaluation set. The core comparisons share the same query identifiers across methods. The benchmark query collections were also used during development; in particular, all 744 Cohere evaluation queries appeared in earlier development experiments, so we also evaluate held-out queries (Sections~\ref{sec:ablation} and~\ref{sec:system}). The four-dataset mechanism experiment uses 200K subsets and independently recomputes their full ground truth. Sources include BGE-M3~\cite{chen2024bgem3}, Wolt-CLIP~\cite{woltclip2024}, and RedCaps~\cite{desai2021redcaps}.

\paragraph{Parameters and repetition}
QuIVer uses $R=64$, construction width $e_c=128$, and the selection rule in Equation~\eqref{eq:prune}. Each main dataset has one constructed graph; Section~\ref{sec:system} reports independent rebuilds. Search widths are selected from
$\{32,48,64,88,128,192,256,384,512,768,1024\}$.
For each of the 90\%, 95\%, and 98\% validation-recall targets, we select the smallest qualifying width. We retain a highest-budget point and additional widths chosen by a fixed budget-spacing rule, up to five distinct widths. Multithread measurements reuse the selected target widths. External methods retain their corresponding validation-selected construction and search configurations.

We measure five throughput rounds after warming the query working set and report their arithmetic mean. A QuIVer round lasts at least five seconds and contains complete passes over the evaluation queries. Latency is collected separately over five complete passes, measuring native service time without client queueing. P95/P99 confidence intervals resample queries as clusters, keeping their repeated measurements together. Recall is the mean fraction of the exact top ten returned. Recall--throughput evaluation follows the benchmarking perspective of ANN-Benchmarks and comparative studies~\cite{aumuller2020annbenchmarks,aumuller2023trends,li2019annanalysis}.

\paragraph{Baselines}
The comparison archive contains 30 external modes: full-precision HNSW implementations in FAISS, HNSWlib, USearch, and VSAG; DiskANN in FP32 and PQ variants; Lucene BBQ; centered and uncentered FAISS BinaryHNSW with FP32 reranking; SVS LVQ4, LVQ8, and two-level LVQ; SymphonyQG; NGT-QG; HSCG; MIRAGE-ANNS; $\alpha$-CNG; and IVF-Flat, PQ, OPQ/FastScan, and RaBitQ variants. FAISS RaBitQ with SQ8 reranking and variants with FP32 reranking are distinct modes. Original-vector use and quantized refinement remain part of each method's measured configuration.

\begin{figure*}[!t]
\centering
\includegraphics[width=\textwidth]{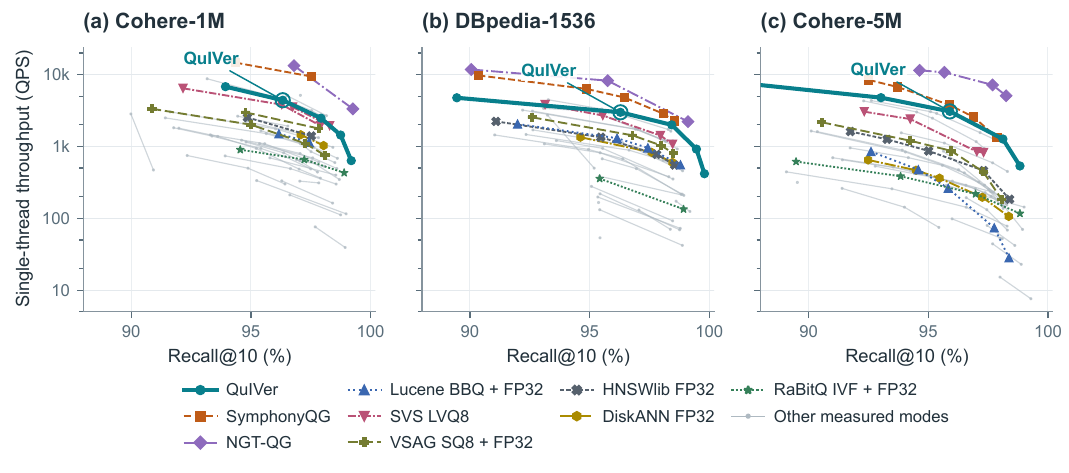}
\caption{Single-thread curves at 88--100\% evaluation recall, with shared logarithmic QPS scales. Colors, markers, and line styles identify representative methods; gray traces retain other measured modes. Lines join points from the same index. Outlined QuIVer markers identify frozen validation-selected 95\% configurations; their horizontal positions show actual evaluation recall.}
\label{fig:curves}
\end{figure*}

Figure~\ref{fig:curves} shows the available single-thread curves, each joining search widths of one constructed index. Table~\ref{tab:main} reports the validation-selected 95\% configurations of nine methods, including the strongest high-throughput quantized graphs and compact alternatives, with their achieved evaluation recall.

For these configurations, SymphonyQG uses degree/beam settings 64/400 on Cohere-1M and 32/200 on the other two workloads, with three refinement iterations. Lucene BBQ uses connectivity/construction-width settings 32/400 on Cohere-1M and 16/200 on the others. SVS LVQ8 and VSAG SQ8 use degree/beam settings 32/200, with SVS padding parameter 32. HNSWlib uses $M=16$ and \texttt{efConstruction}=200. DiskANN uses degree/beam settings 32/200 on Cohere-1M and DBpedia-1536 and 64/400 on Cohere-5M. NGT-QG uses ANNG degree 40, outgoing degree 32, and quantizer parameters $Q=2$, $k=32$, $p=8$. RaBitQ IVF uses 4,096 inverted lists and \texttt{nprobe}=128 on all three workloads in Table~\ref{tab:main}, with its official 32-bit FP32-refinement mode. The companion data retain each construction configuration and its validation-selected search parameters.

\paragraph{Measurement definitions}
$T_{\mathrm{idx}}$ is the recorded index preparation time. For QuIVer it covers graph construction and SM installation. For SymphonyQG, HNSWlib, VSAG, and RaBitQ it is the official build call, and for SVS the LVQ quantization and Vamana build calls; serialization is excluded for these methods and for QuIVer. NGT-QG's build commands include their internal persistence, Lucene's time spans insertion through forceMerge and commit, and DiskANN's includes its PQ training.

$M_{\mathrm{nav}}$ sums the persistent navigation components, including any full-precision navigation data. $M_{\mathrm{total}}$ adds the separate original-vector store used for reranking, which methods such as SymphonyQG and SVS do not need. Query-process RSS is retained in the companion measurement data.

\begin{table*}[!t]
\caption{Validation-selected 95\% operating points. Recall $r$ is the observed evaluation Recall@10; QPS values average five rounds. $T_{\mathrm{idx}}$ includes QuIVer's graph construction and SM installation. Memory columns are logical component subtotals in GiB; $M_{\mathrm{total}}-M_{\mathrm{nav}}$ is the separate original-vector store. P99 is single-thread service latency in milliseconds; \missing{} indicates unavailable native latency.}
\label{tab:main}
\centering
\footnotesize
\setlength{\tabcolsep}{3.5pt}
\begin{quivercheckedtable}{main-comparison}
\begin{tabular}{llrrrrrrrrr}
\toprule
Dataset & Method & $r$ (\%) & 1T QPS & 8T QPS & 16T QPS & 32T QPS & $T_{\mathrm{idx}}$ (s) & $M_{\mathrm{nav}}$ & $M_{\mathrm{total}}$ & P99 \\
\midrule
Cohere-1M & \textbf{QuIVer} & 96.33 & 4,396 & 31,061 & 48,357 & 63,103 & 100.6 & 0.920 & 3.781 & 0.391 \\
Cohere-1M & SymphonyQG & 94.35 & 14,672 & 90,243 & 138,781 & 168,842 & 585.3 & 11.444 & 11.444 & 0.121 \\
Cohere-1M & NGT-QG & 96.80 & 13,296 & 71,535 & 79,122 & 79,138 & 1,703.6 & 6.021 & 8.882 & 0.232 \\
Cohere-1M & Lucene BBQ + FP32 & 96.17 & 1,503 & 12,864 & 22,519 & 30,437 & 3,886.1 & 0.171 & 3.032 & 1.043 \\
Cohere-1M & SVS LVQ8 & 96.29 & 3,814 & 25,527 & 39,183 & 45,320 & 260.8 & 0.868 & 0.868 & \missing \\
Cohere-1M & VSAG SQ8 + FP32 & 95.00 & 2,010 & 13,128 & 15,933 & 21,507 & 352.4 & 0.859 & 3.765 & 0.717 \\
Cohere-1M & HNSWlib FP32 & 94.85 & 2,532 & 10,149 & 10,439 & 10,539 & 261.2 & 3.007 & 3.007 & 0.638 \\
Cohere-1M & DiskANN FP32 & 97.08 & 1,455 & 7,246 & 7,843 & 7,791 & 371.0 & 3.040 & 3.040 & 1.102 \\
Cohere-1M & RaBitQ IVF + FP32 & 97.23 & 657 & 4,872 & 8,423 & 12,972 & 58.2 & 0.123 & 2.984 & 2.011 \\
\addlinespace[3pt]
DBpedia-1536 & \textbf{QuIVer} & 96.29 & 2,987 & 20,272 & 28,926 & 31,138 & 149.3 & 1.265 & 6.930 & 0.660 \\
DBpedia-1536 & SymphonyQG & 96.45 & 4,803 & 28,437 & 38,906 & 40,846 & 521.7 & 13.690 & 13.690 & 0.304 \\
DBpedia-1536 & NGT-QG & 95.76 & 8,229 & 38,429 & 41,231 & 39,338 & 3,991.5 & 11.774 & 17.439 & 0.294 \\
DBpedia-1536 & Lucene BBQ + FP32 & 96.15 & 1,303 & 8,421 & 18,519 & 28,581 & 3,765.3 & 0.266 & 5.931 & 1.157 \\
DBpedia-1536 & SVS LVQ8 & 95.55 & 2,692 & 16,502 & 20,868 & 21,547 & 512.9 & 1.567 & 1.567 & \missing \\
DBpedia-1536 & VSAG SQ8 + FP32 & 96.79 & 1,427 & 9,826 & 13,064 & 14,401 & 704.5 & 1.560 & 7.279 & 1.062 \\
DBpedia-1536 & HNSWlib FP32 & 95.47 & 1,345 & 4,872 & 4,949 & 4,934 & 531.2 & 5.809 & 5.809 & 1.088 \\
DBpedia-1536 & DiskANN FP32 & 97.59 & 828 & 3,427 & 3,697 & 3,521 & 754.7 & 5.842 & 5.842 & 1.776 \\
DBpedia-1536 & RaBitQ IVF + FP32 & 95.42 & 357 & 2,616 & 4,214 & 5,176 & 93.2 & 0.227 & 5.892 & 3.606 \\
\addlinespace[3pt]
Cohere-5M & \textbf{QuIVer} & 95.90 & 3,075 & 21,167 & 31,557 & 34,109 & 524.6 & 4.601 & 18.906 & 0.542 \\
Cohere-5M & SymphonyQG & 95.87 & 3,801 & 21,593 & 31,701 & 35,276 & 1,130.2 & 35.763 & 35.763 & 0.394 \\
Cohere-5M & NGT-QG & 95.67 & 10,751 & 56,286 & 62,269 & 58,039 & 15,774.1 & 30.135 & 44.441 & 0.320 \\
Cohere-5M & Lucene BBQ + FP32 & 95.83 & 262 & 1,945 & 3,469 & 5,010 & 7,601.2 & 0.757 & 15.062 & 6.436 \\
Cohere-5M & SVS LVQ8 & 94.25 & 2,406 & 16,046 & 21,182 & 21,031 & 1,357.7 & 4.340 & 4.340 & \missing \\
Cohere-5M & VSAG SQ8 + FP32 & 95.95 & 869 & 5,744 & 8,581 & 9,173 & 1,141.8 & 4.240 & 18.583 & 1.801 \\
Cohere-5M & HNSWlib FP32 & 95.01 & 866 & 3,224 & 3,325 & 3,323 & 1,007.2 & 15.034 & 15.034 & 1.568 \\
Cohere-5M & DiskANN FP32 & 95.47 & 362 & 1,814 & 1,794 & 1,884 & 5,640.3 & 16.093 & 16.093 & 3.781 \\
Cohere-5M & RaBitQ IVF + FP32 & 96.98 & 218 & 1,359 & 2,047 & 2,402 & 257.3 & 0.540 & 14.845 & 7.551 \\
\addlinespace[3pt] 
\bottomrule
\end{tabular}
\end{quivercheckedtable}
\end{table*}

\subsection{Pareto Efficiency across Construction, Memory, and Throughput}
\label{sec:mainresults}
\begin{figure*}[!t]
\centering
\includegraphics[width=\textwidth]{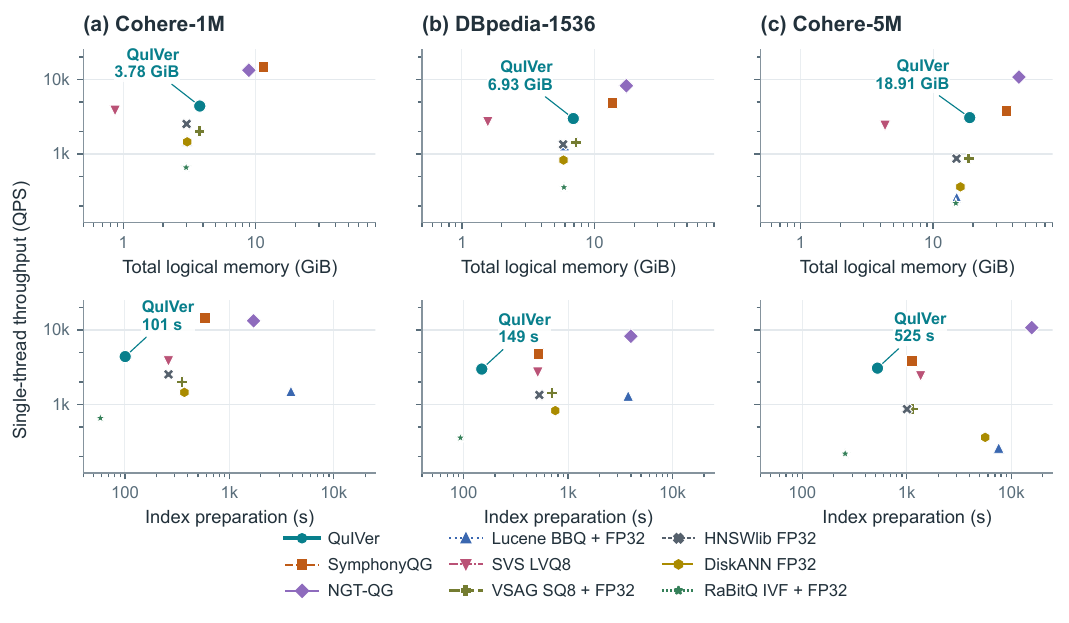}
\caption{System trade-offs at the frozen configurations of Table~\ref{tab:main}: total logical memory $M_{\mathrm{total}}$ versus QPS (upper), and recorded index preparation $T_{\mathrm{idx}}$ versus QPS (lower). Logarithmic scales are shared across datasets. Upper-left positions indicate lower cost and higher throughput. The table supplies achieved recall, navigation memory, and timing definitions.}
\label{fig:memory}
\end{figure*}

QuIVer's validation-selected 95\% configurations lie on the measured Pareto frontiers of all three core workloads at both one and 32 threads. We maximize achieved recall and throughput and minimize $T_{\mathrm{idx}}$ and memory: no measured point matches or improves every dimension and improves at least one. This holds whether memory is $M_{\mathrm{total}}$, $M_{\mathrm{nav}}$, or, for points with recorded RSS, query-process RSS. The comparison uses the full measured curves, including the actual recalls of points that miss their validation targets. Figure~\ref{fig:memory} shows the cost--throughput projections.

\paragraph{Cohere-1M}
At 96.33\% recall, QuIVer reaches 4,396 single-thread QPS. Lucene BBQ reaches 1,503 QPS at 96.17\%, and SVS LVQ8 reaches 3,814 at 96.29\%. These close-recall points give throughput ratios of $2.92\times$ and $1.15\times$, respectively. QuIVer's graph build takes 89.40 s and SM installation 11.16 s. Its total logical memory is 3.781 GiB, against 3.032 GiB for BBQ and 0.868 GiB for LVQ8, whose compressed vectors need no separate store.

QuIVer uses $3.03\times$ and $2.35\times$ less total logical memory than SymphonyQG and NGT-QG ($12.44\times$ and $6.54\times$ less navigation storage), and its recorded preparation is $5.82\times$ and $16.94\times$ faster.

\paragraph{DBpedia-1536}
At 96.29\% recall, QuIVer reaches 2,987 QPS with 1.265 GiB of navigation components and 6.930 GiB in total. Lucene BBQ reaches 1,303 at 96.15\%, while SVS LVQ8 reaches 2,692 at 95.55\%. QuIVer's 128.45 s graph build and 20.83 s installation together take 149.28 s. Relative to LVQ8, this gives higher recall, $1.11\times$ throughput, and $3.44\times$ faster preparation, whereas LVQ8 needs $4.42\times$ less total memory.

Table~\ref{tab:high} reports the 98\% target configurations. QuIVer reaches 98.43\% at 1,997 QPS; LVQ8 reaches 98.47\% at 1,073 QPS, and SymphonyQG reaches 98.56\% at 2,265 QPS. At these nearby recalls, QuIVer delivers $1.86\times$ LVQ8's throughput and 88\% of SymphonyQG's throughput with $1.98\times$ less total logical memory than SymphonyQG. The corresponding query-process RSS is approximately 7.00 GiB for QuIVer, 13.85 GiB for SymphonyQG, and 1.73 GiB for LVQ8.

\paragraph{Cohere-5M}
The five-million-vector workload combines near-SymphonyQG concurrent throughput with lower memory and preparation cost. QuIVer reaches 95.90\% recall at 3,075 single-thread QPS; SymphonyQG reaches 95.87\% at 3,801 QPS. QuIVer retains 81\% of its single-thread throughput and 97\% of its 32-thread throughput (34,109 versus 35,276 QPS). Total logical memory falls from 35.763 to 18.906 GiB (47.1\%) and navigation storage to 4.601 GiB (87.1\%); preparation is $2.15\times$ faster. SymphonyQG's payload comprises 14.305 GiB of FP32 navigation vectors, 19.073 GiB of per-edge local codes, 1.788 GiB of edge factors, and 0.596 GiB of neighbor identifiers.

QuIVer also delivers $11.74\times$ Lucene BBQ's single-thread throughput at nearby recall (95.83\% for BBQ), using $1.26\times$ its 15.062 GiB total logical memory. NGT-QG reaches 10,751 QPS at 95.67\% recall with 44.441 GiB in total and a 15,774 s recorded build. QuIVer's 465.22 s graph build and 59.37 s SM installation give $30.07\times$ faster preparation and $2.35\times$ less total logical memory.

\paragraph{Uncertainty and matched recall}
Query-bootstrap 95\% intervals for QuIVer's recall are [95.34, 97.20]\%, [96.03, 96.55]\%, and [94.88, 96.87]\% on the three workloads, and its five-round single-thread throughput varies by at most 1.9\% (CV). Paired with SymphonyQG, the recall differences on DBpedia-1536 ($-0.16$ pp) and Cohere-5M ($+0.03$ pp) have intervals containing zero. Interpolating each fixed configuration's single-thread curve to 95\% recall gives QuIVer 5,599, 3,261, and 3,528 QPS. SymphonyQG reaches $2.39\times$, $1.89\times$, and $1.36\times$ these values and NGT-QG $2.65\times$ and $3.18\times$ on the last two workloads, whereas QuIVer reaches $1.25\times$, $1.12\times$, and $1.95\times$ LVQ8's. QuIVer's frontier position thus combines throughput with preparation time and memory rather than maximizing throughput alone.

\begin{table}[t]
\caption{Validation-selected 98\% configurations. Each dataset reports achieved evaluation recall (\%) and single-thread QPS.}
\label{tab:high}
\centering
\footnotesize
\setlength{\tabcolsep}{1.4pt}
\begin{quivercheckedtable}{high-recall}
\begin{tabular}{lrrrrrr}
\toprule
& \multicolumn{2}{c}{\shortstack{Cohere\\1M}} & \multicolumn{2}{c}{\shortstack{DBpedia\\1536}} & \multicolumn{2}{c}{\shortstack{Cohere\\5M}}\\
Method & $r$ & QPS & $r$ & QPS & $r$ & QPS\\
\midrule
QuIVer & 98.75 & 1,441 & 98.43 & 1,997 & 98.12 & 1,269 \\
SymphonyQG & 97.54 & 9,379 & 98.56 & 2,265 & 97.88 & 1,323 \\
NGT-QG & 99.26 & 3,333 & 99.11 & 2,216 & 98.25 & 5,072 \\
Lucene BBQ + FP32 & 96.17 & 1,503 & 98.80 & 556 & 98.37 & 28 \\
SVS LVQ8 & 98.29 & 1,943 & 98.47 & 1,073 & 97.31 & 823 \\
HNSWlib FP32 & 97.53 & 1,399 & 98.55 & 550 & 98.40 & 184 \\ 
\bottomrule
\end{tabular}
\end{quivercheckedtable}
\end{table}

\subsection{Concurrent Queries and Tail Latency}
\label{sec:threads}
Table~\ref{tab:main} reports throughput at 1--32 threads on the same graphs and search parameters. At the 95\% operating points, QuIVer's 32-thread speedups over its own single-thread throughput are $14.35\times$, $10.42\times$, and $11.09\times$ on Cohere-1M, DBpedia-1536, and Cohere-5M.

On Cohere-5M, SymphonyQG reaches a $9.28\times$ speedup at 32 threads, BBQ $19.13\times$, and HNSWlib $3.84\times$. Their absolute throughputs are 35,276, 5,010, and 3,323 QPS, respectively; QuIVer reaches 34,109 QPS.

Single-thread P99 at the 95\% configurations is 0.391 ms, 0.660 ms, and 0.542 ms for QuIVer on the three core datasets. On Cohere-5M, BBQ and SymphonyQG have P99 values of 6.436 ms and 0.394 ms. These service-time percentiles come from the separate latency passes.

\subsection{Representation and Scale}
\label{sec:scope}
Table~\ref{tab:datasets} fixes the largest measured width, $e=1024$, and the 64-candidate FP32 budget across datasets. Search quality depends on embedding geometry and on how the retrieval task relates queries to the database.

High recall extends across several semantic collections. MiniLM, BGE-M3, both DBpedia dimensionalities, Arxiv-Nomic, CCNews-Nomic, CodeSearch-Jina, and GooAQ-DistilRoBERTa all exceed 98.9\%. Cohere-5M and MSMARCO-5M reach 98.83\% and 98.94\%. The 3,072-dimensional DBpedia representation reaches 99.81\% with a 1.973 GiB navigation subtotal. The reported index stages for MSMARCO-5M take 413.4 s, with 5.197 GiB of navigation components.

The learned representations cover different training objectives and retrieval tasks. GloVe's static word vectors, trained from word co-occurrence~\cite{pennington2014glove}, reach 74.52\% recall. Wolt-CLIP and RedCaps reach 84.68\% and 95.03\%, respectively; our RedCaps evaluation uses the released \texttt{random\_test} query group. COCO-Nomic uses text queries against an image corpus, whereas Landmark-Nomic uses image queries against an image corpus generated by the same Nomic vision encoder~\cite{jaasaari2025vibe}. Their recalls are 25.05\% and 98.02\%, respectively. These workloads differ in both corpus content and the relationship between query and database distributions. Landmark-DINO reaches 98.67\%.

Among the 768-dimensional synthetic collections, Random-Sphere reaches 15.58\% and the low-rank collection 93.23\%. Native-L2 SIFT and GIST reach 28.39\% and 3.28\%; SIFT's coordinates are non-negative, so its sign plane carries little or no information. Section~\ref{sec:mechanism} examines representation effects through controlled changes to graph construction and query encoding, measuring graph quality and candidate losses.

\begin{table*}[!t]
\caption{The 21-dataset QuIVer evaluation at the fixed largest width $e=1024$ and FP32 budget 64. Queries are evaluation queries after removing the 256-query tuning split. $T_{\mathrm{idx}}$ is graph construction plus SM installation; $M_{\mathrm{nav}}$ is the persistent navigation-component subtotal. Only the first three datasets have the external comparisons in Table~\ref{tab:main}.}
\label{tab:datasets}
\centering
\footnotesize
\setlength{\tabcolsep}{5pt}
\begin{quivercheckedtable}{datasets}
\begin{tabular}{lrrrlrrrr}
\toprule
Dataset & Vectors & $d$ & Queries & Metric & Recall (\%) & 1T QPS & $T_{\mathrm{idx}}$ (s) & $M_{\mathrm{nav}}$ (GiB)\\
\midrule
Cohere-1M & 1,000,000 & 768 & 744 & Cosine & 99.19 & 635 & 100.6 & 0.920 \\
DBpedia-1536 & 990,000 & 1536 & 9,744 & Cosine & 99.80 & 418 & 149.3 & 1.265 \\
Cohere-5M & 5,000,000 & 768 & 744 & Cosine & 98.83 & 535 & 524.6 & 4.601 \\
MiniLM & 1,000,000 & 384 & 744 & Cosine & 99.49 & 690 & 81.2 & 0.712 \\
Wolt-CLIP & 1,000,000 & 512 & 744 & Cosine & 84.68 & 924 & 65.6 & 0.801 \\
BGE-M3 & 1,000,000 & 1024 & 744 & Cosine & 99.99 & 684 & 91.3 & 1.039 \\
DBpedia-3072 & 990,000 & 3072 & 9,744 & Cosine & 99.81 & 246 & 261.6 & 1.973 \\
RedCaps-1M & 1,000,000 & 512 & 9,744 & Cosine & 95.03 & 819 & 57.9 & 0.801 \\
GloVe & 1,183,514 & 100 & 9,744 & Cosine & 74.52 & 435 & 68.3 & 0.701 \\
SIFT & 1,000,000 & 128 & 9,744 & L2 & 28.39 & 549 & 50.9 & 0.592 \\
GIST & 1,000,000 & 960 & 744 & L2 & 3.28 & 1,478 & 64.3 & 1.024 \\
Random-Sphere & 1,000,000 & 768 & 744 & Cosine & 15.58 & 287 & 164.7 & 0.920 \\
Synthetic-LR & 1,000,000 & 768 & 744 & Cosine & 93.23 & 400 & 135.3 & 0.920 \\
MSMARCO-5M & 5,000,000 & 1024 & 9,744 & Cosine & 98.94 & 610 & 413.4 & 5.197 \\
Arxiv-Nomic & 1,000,000 & 768 & 744 & Cosine & 99.53 & 713 & 92.1 & 0.920 \\
CCNews-Nomic & 495,328 & 768 & 744 & Cosine & 98.92 & 783 & 39.8 & 0.456 \\
COCO-Nomic & 282,360 & 768 & 744 & Cosine & 25.05 & 781 & 18.9 & 0.260 \\
CodeSearch-Jina & 1,000,000 & 768 & 744 & Cosine & 99.62 & 671 & 95.9 & 0.920 \\
GooAQ-DistilRoBERTa & 1,000,000 & 768 & 744 & Cosine & 99.93 & 750 & 88.6 & 0.920 \\
Landmark-Nomic & 760,757 & 768 & 744 & Cosine & 98.02 & 968 & 58.7 & 0.700 \\
Landmark-DINO & 760,757 & 768 & 744 & Cosine & 98.67 & 826 & 65.1 & 0.700 \\ 
\bottomrule
\end{tabular}
\end{quivercheckedtable}
\end{table*}

\subsection{Scoring, Retention, and Exact Budget}
\label{sec:ablation}
We use a Cohere-200K dataset to compare scoring rules under a common graph and search implementation. The query pool is divided into 384 calibration queries for the optional correction model, 128 validation queries, and 488 evaluation queries. The production SM-L2 rule requires no calibration queries.

\paragraph{Strong untrained scoring controls}
Table~\ref{tab:scoring} compares fixed-amplitude BQ, reconstruction inner product, reconstruction cosine, reconstruction L2, and an optional shared correction model. The correction model adjusts the reconstructed inner product using a query-specific error observed at an exactly scored anchor and a precomputed per-vertex coefficient; its anchor consumes part of the FP32 budget.

At width 64, unnormalized inner product achieves only 76.72\% recall. Reconstruction cosine and L2 reach 97.13\% and 96.48\%, compared with 96.31\% for shared correction. At validation-selected high-recall points, L2 reaches 98.01\% at 4,260 QPS and correction reaches 98.28\% at 4,269 QPS. A second common engine compares the cosine variants over widths 64--512 on 744 queries: normalizing online or folding the norm into the amplitudes offline yields identical recall and throughput within 5\%. Both trail SM-L2 by 0.46--0.62 pp on Cohere-200K at $e\ge256$; on Wolt-200K they lead by 1.94 pp at $e=64$ and are statistically indistinguishable at larger widths. These results support the training-free L2 production path.

\begin{table}[t]
\caption{Cohere-200K scoring controls at $e=64$ and 64 total exact evaluations, including one anchor for every row in this controlled comparison. Production SM-L2 omits the anchor. Throughput belongs to the common ablation engine.}
\label{tab:scoring}
\centering
\begin{quivercheckedtable}{scoring}
\begin{tabular}{lrrr}
\toprule
Score & Recall (\%) & 1T QPS & QPS CV (\%)\\
\midrule
Fixed-amplitude BQ & 95.06 & 6,974 & 2.03\\
Reconstructed inner product & 76.72 & 6,663 & 3.73\\
Reconstructed cosine & 97.13 & 6,139 & 2.95\\
Reconstructed L2 & 96.48 & 6,897 & 0.72\\
Shared correction & 96.31 & 7,124 & 1.84\\
\bottomrule
\end{tabular}
\end{quivercheckedtable}
\end{table}

\paragraph{Where are the neighbors lost?}
The oracle decomposition in Equation~\eqref{eq:decomposition} distinguishes candidate exposure from retention. At width 64, the unnormalized inner-product search visits 96.31\% of the true neighbors but retains only 76.93\% in its final beam. Reconstruction L2 preserves useful candidates already exposed by traversal and reaches 96.48\% final recall (Table~\ref{tab:scoring}). Applying shared correction only to the final ordering raises recall from 76.72\% to 76.93\%, whereas applying it during navigation and beam maintenance reaches 96.27\%.

\paragraph{Exact reranking budget}
At width 128, fixed-amplitude BQ improves from 95.70\% to 98.20\% when the total exact budget increases from 64 to 128. Its plateau at budget 64 is therefore partly a reranking-selection bottleneck. For production-style L2 without an anchor, budgets 16, 32, and 64 yield 88.20\%, 95.84\%, and 98.05\% recall at width 128. Shared correction, charging its anchor to the same budgets, stays within one percentage point and gains about 3\% throughput at the 95\% and 98\% points; the untrained path avoids its calibration state and per-query anchor. With the score fixed, layout, prefetching, and grouped scoring together improve throughput by approximately 25\%.

\paragraph{Held-out queries}
To test whether these choices depend on development queries, we draw 1,024 document vectors that were never used in development from 32 source blocks of the public 35M-vector Cohere corpus, recompute exact Cohere-200K ground truth, and resample source blocks for intervals. SM-L2 exceeds both cosine variants by 0.82--2.08 pp at widths 64--512, and the construction and representation variants evaluated on these queries keep their relative order (Section~\ref{sec:mechanism}). Absolute recall at the same width is 0.67--1.26 pp below that of the benchmark queries.

\subsection{Construction--Query Crossovers and Geometry}
\label{sec:mechanism}
We construct four controlled graphs per 200K dataset: standard BQ, FP32, centered BQ, and per-dimension-threshold BQ. A common controlled builder starts from a random graph with 64 out-edges per vertex and uses a fixed insertion order and seed, degree bound 64, construction width 128, and two passes with $\alpha=1.0$ and $1.2$. Each graph discovers and prunes candidates with its own score $D$, either $D_{\mathrm{BQ}}$ or squared Euclidean distance without offset, and rejects $y$ when $\alpha D(y,z)\le D(x,y)$, the standard Vamana direction rather than Equation~\eqref{eq:prune}. These builds isolate representation effects under one rule; Section~\ref{sec:system} measures production construction.

Table~\ref{tab:cross} crosses graph construction with SM-L2 or exact navigation, at width 128 and reranking budget 64. Cohere reaches 97.67\% with BQ construction and SM navigation, versus 99.53\% with FP32 in both stages. On Wolt, replacing only the BQ graph with the FP32 graph improves SM recall from 94.37\% to 98.63\%. Graph quality remains a substantial factor even when the query representation is unchanged.

The corresponding BQ/FP32 graph-construction times are 137.64/477.81 s on Cohere, 91.77/236.06 s on Wolt, 183.56/704.17 s on Random-Sphere, and 91.83/97.13 s on SIFT. These times cover candidate discovery, score-based selection, and graph updates in the common controlled builder.

\begin{table}[t]
\caption{Controlled 200K graph/query precision crossovers: Recall@10 (\%), $e=128$, final budget 64. Column headings specify construction and navigation precision.}
\label{tab:cross}
\centering
\small
\setlength{\tabcolsep}{3pt}
\begin{quivercheckedtable}{crossovers}
\begin{tabular}{lrrrr}
\toprule
& BQ/SM & BQ/FP32 & FP32/SM & FP32/FP32\\
\midrule
Cohere & 97.67 & 98.59 & 98.51 & 99.53\\
Wolt-CLIP & 94.37 & 96.67 & 98.63 & 99.97\\
Random-Sphere & 11.47 & 12.10 & 11.69 & 13.75\\
SIFT, native L2 & 31.71 & 61.63 & 43.33 & 99.99\\
\bottomrule
\end{tabular}
\end{quivercheckedtable}
\end{table}

SIFT separates graph reachability from approximate retention particularly clearly. With the FP32 graph and SM navigation, the visited set covers 97.54\% of the true neighbors. Only 57.14\% remain in the candidate pool and 43.33\% reach final reranking. The FP32 graph alone cannot recover accuracy when the low-bit query score discards the relevant candidates. Conversely, exact query navigation on a BQ graph reaches only 61.63\%, exposing construction loss as well.

Centering both construction and query codes improves recall to 98.13\%, 98.09\%, and 73.21\% on Cohere, Wolt, and SIFT, whereas Random-Sphere stays near its uncentered BQ/SM value (11.44\% versus 11.47\%). Per-dimension thresholds yield 98.09\%, 95.30\%, 11.06\%, and 31.81\% on the four datasets. Both transformations store $d$ database-derived FP32 values. The main study uses one corpus-independent, per-vector encoding across datasets. These variants quantify gains from adapting to collection geometry.

\begin{table}[t]
\caption{Selection scores and rejection rules in a common 200K engine with Hamming discovery. Each rule uses its score for ordering and rejection. Columns report final edges from backfill (BF), graph-construction time, and single-thread Recall@10 (\%) at $e=64/128$.}
\label{tab:build}
\centering
\footnotesize
\setlength{\tabcolsep}{2.4pt}
\begin{quivercheckedtable}{selection-rules}
\begin{tabular}{lrrrrrr}
\toprule
& \multicolumn{3}{c}{Cohere-200K} & \multicolumn{3}{c}{Wolt-200K}\\
\cmidrule(lr){2-4}\cmidrule(lr){5-7}
Rule & BF (\%) & s & $r_{64}/r_{128}$ & BF (\%) & s & $r_{64}/r_{128}$\\
\midrule
Eq.~\eqref{eq:prune}, $\alpha=1.2$ & 98 & 12.7 & 96.1/98.1 & 98 & 8.4 & 89.0/93.7\\
Eq.~\eqref{eq:prune}, $\alpha=1.0$ & 63 & 15.9 & 97.4/98.7 & 71 & 11.2 & 97.3/98.4\\
Standard, $D_{\mathrm{BQ}}$ & 0 & 17.9 & 96.1/98.2 & 0 & 15.3 & 93.4/96.6\\
Standard, integer L2 & 1 & 21.1 & 97.0/98.4 & 12 & 20.0 & 97.0/97.6\\
\bottomrule
\end{tabular}
\end{quivercheckedtable}
\end{table}

\paragraph{Selection rule}
Table~\ref{tab:build} compares selection scores and rejection rules under common Hamming discovery. Under Equation~\eqref{eq:prune}, 98\% of the final edges again come from backfill. Admitting more diverse edges, by setting $\alpha=1$, applying $\alpha$ in the standard direction, or using the integer squared distance $\|u_x-u_y\|^2$ for ordering and rejection, raises Wolt recall at $e=64$ by 4.4--8.3 pp but lengthens graph construction by 25--138\%. On Cohere the gain is at most 1.4 pp. At matched recall, $\alpha=1$ delivers $3.46\times$ the evaluated rule's single-thread throughput on Wolt and $1.14\times$ on Cohere (log-linear interpolation). The main study fixes $\alpha=1.2$ for rapid preparation across datasets. The $\alpha=1$ variant supplies a higher-query-throughput operating point in this 200K control. These variants trade construction time for graph quality on workloads such as Wolt-CLIP, where Table~\ref{tab:cross} shows that the graph limits recall.

\paragraph{Same-budget scalar quantization}
With a common reconstructed-L2 builder and search ($e=128$, budget 64), the SM representation reaches 97.53\%, 97.92\%, and 39.93\% on Cohere, Wolt, and SIFT. A standard per-vector min--max 2-bit quantizer reaches 6.52\%, 28.33\%, and 91.42\%; on Cohere, 95.5\% of its coordinates fall into a single level. Subtracting the database mean first, as LVQ does~\cite{aguerrebere2023lvq}, gives 82.33\%, 69.87\%, and 71.28\%. Its losses concentrate in final selection: at $e=512$ on Cohere, its beam retains 94.38\% of the true neighbors, but its 64 best-scored candidates retain 84.37\%. At the same bit budget, the sign--magnitude split thus outperforms uniform levels on Cohere and Wolt, whereas uniform levels suit non-negative SIFT.

\paragraph{Ordering and candidate sources}
Independent 200K controls separate ordering from rejection scoring, keeping Equation~\eqref{eq:prune}'s direction and $\alpha=1.2$. Two seeds use SM-L2 queries at $e=128$, budget 64, and five rounds over all 744 queries. Integer L2 ordering alone changes recall by $-0.17/+0.22$ pp on Cohere/Wolt. Using it for both ordering and rejection changes recall by $-0.16/+0.88$ pp, with both Wolt paired 95\% intervals spanning zero. Removing random supplementation reduces graph time by 6.3\%/4.2\% with recall changes of $-0.03/+0.47$ pp. Disabling Hamming discovery while retaining 128 random draws yields 12.12\%/12.55\% recall.

\subsection{Construction Parallelism, Rebuilds, and SSD Mode}
\label{sec:system}
On Cohere-1M, deferred reverse-edge pruning takes 603.48 s with one construction thread, 89.53 s with eight, 53.71 s with sixteen, and 33.91 s with thirty-two. The eight-thread value averages three builds, ranging from 89.09 to 89.90 s. Eager pruning at eight threads averages 336.05 s across three builds. Thus, deferring repeated pruning reduces the measured eight-thread construction time by $3.75\times$. At fixed width 128, the corresponding recall ranges are 97.193--97.234\% for deferred pruning and 96.783--96.803\% for eager pruning. This comparison measures the implemented construction policies, including their effect on the final graph.

Three independent rebuilds of the frozen Cohere-1M configuration reproduce the operating points of Tables~\ref{tab:main} and~\ref{tab:high}: Recall@10 is 96.30--96.45\% at $e=88$ (96.33\% in Table~\ref{tab:main}) and 98.75--98.78\% at $e=384$ (98.75\%). Across rebuilds, graph construction time varies by 0.19\% and throughput at 1--32 threads by 0.21--0.94\% (coefficients of variation).

\paragraph{Population-count kernel}
Caching strong-bit counts reduces construction scoring to four population-count reductions. Equal-layout microbenchmarks reduce scoring time by 19.2\%/39.0\% on Cohere/Wolt. Three paired 200K TDB builds change mean construction time from 12.256/7.538 s to 12.151/7.446 s, with four cache bytes per vector. Both paired 95\% timing intervals span zero. On 4,096-node single-thread builds, serialized indices, labels, and score bits match the original kernel.

\begin{table}[t]
\caption{SSD mode on held-out query streams at $e=88$ with 64 INT8 candidates and a 128 MiB cache. Values are arithmetic means over five streams, including the per-stream P99 values. Reads use \texttt{O\_DIRECT}. RSS is steady-state query-process RSS. Recall@10 is 88.97\% at 1M and 86.59\% at 5M.}
\label{tab:ssd}
\centering
\footnotesize
\setlength{\tabcolsep}{3pt}
\begin{quivercheckedtable}{ssd}
\begin{tabular}{lrrrrr}
\toprule
Data & CPUs & QPS & P99 (ms) & KiB/query & RSS (GiB)\\
\midrule
Cohere-1M & 1 & 1,656 & 0.87 & 17.7 & 1.2\\
Cohere-5M & 1 & 1,120 & 1.82 & 41.1 & 5.0\\
Cohere-1M & 32 & 10,258 & 4.73 & 17.6 & 1.2\\
Cohere-5M & 32 & 4,368 & 20.00 & 41.0 & 5.0\\
\bottomrule
\end{tabular}
\end{quivercheckedtable}
\end{table}

\paragraph{SSD mode}
Table~\ref{tab:ssd} evaluates the SSD mode of Section~\ref{sec:ssd} on streams that never repeat a query: five 20K-query streams of held-out Cohere document vectors, each starting from an empty cache and measured over its last 10K queries. Query and I/O threads share the stated CPUs. Each query reads at most 64 KiB, and the cache cuts this to 17.7 KiB on Cohere-1M. Without a long-term cache, throughput falls by only 8--12\% from 1M to 5M vectors. With it, the cache hit rate falls from 72\% to 36\%; at 32 CPUs, throughput times bytes read stays near 175 MiB/s at both scales, so the slowdown follows the added reads. Steady-state RSS is 1.2 and 5.0 GiB, against 3.8 and 19.1 GiB in memory.

INT8 rows cost 1.24 pp on the 744 evaluation queries (95.13\% versus 96.37\% with FP32 rows on the same rebuilt index and width). The held-out stream is a measurably harder workload: its relative contrast RC$_{10}$ is 6.95\% lower and its LID$_{100}$ 62.92\% higher than for the benchmark queries~\cite{he2012difficulty,aumuller2021lid}. On 1,000 pre-registered stream queries, FP32 reranking of the archived candidates reaches 90.59\% at 1M, and INT8 reaches 88.97\% (86.59\% at 5M). Reweighting QuIVer's per-query benchmark recall from Table~\ref{tab:main} to the LID or RC distribution of these queries predicts 86.8\% or 92.5\%, bracketing 90.59\%. On the benchmark queries, the recall of all nine methods declines with LID (Spearman $\rho$ from $-0.18$ to $-0.37$; $-0.28$ for QuIVer).

\section{Related Work}
\label{sec:related}
\paragraph{Graph search and compressed construction}
Small-world routing~\cite{kleinberg2000smallworld} and search in growth-restricted metrics~\cite{karger2002growth} provide foundations for graph-based neighbor search. HNSW~\cite{malkov2020hnsw}, DiskANN/Vamana~\cite{jayaram2019diskann}, and NSG~\cite{fu2019nsg} turn graph navigation into practical indices. Monotonic proximity graphs~\cite{zhu2021monotonic} and error-bounded $\delta$-EMG~\cite{xiang2025emg} study conditions supporting search progress. LVQ~\cite{aguerrebere2023lvq} constructs graphs in a compressed space, and FAISS BinaryHNSW indexes binary vectors under Hamming distance~\cite{faissbinary}.

\paragraph{Indexing and query systems}
Graph-search surveys expose the distinct effects of initialization, candidate discovery, pruning, and query traversal~\cite{wang2021graphsurvey}. Flash uses compact codes and CPU-aware operations to accelerate construction~\cite{wang2025flash}; PiPNN organizes scalable construction around partitions and online pruning~\cite{dong2026pipnn}. SHG adds shortcuts to hierarchical navigation~\cite{gong2025shg}, PAG uses projections to reduce exact comparisons~\cite{lu2026pag}, and VSAG~\cite{zhong2025vsag} and USearch~\cite{usearch2024} address practical deployment.

\paragraph{Quantization and graph co-design}
NGT-QG stores neighboring codes contiguously for batched estimates~\cite{ngt}. SymphonyQG combines quantization, graph refinement, layout, and exact scoring during traversal~\cite{gou2025symphony}. Routing-guided product quantization trains a quantizer to preserve graph neighborhood and routing features~\cite{yue2024rpq}. QuIVer instead derives its codes from per-vector statistics without training.

\paragraph{Vector compression and exact refinement}
Product quantization~\cite{jegou2011pq}, optimized product quantization~\cite{ge2013opq}, and FAISS~\cite{johnson2021faiss} provide a broad range of compression choices, surveyed by Matsui et al.~\cite{matsui2018pqs}. PQ Fast Scan~\cite{andre2015pqfs} and Quick ADC~\cite{andre2017quickadc} show how code layout and SIMD lookup affect query cost. RaBitQ provides theoretically controlled distance estimation at one or more bits~\cite{gao2024rabitq,gao2025extended}. Lucene BBQ uses binary database codes, asymmetric query precision, and correction factors~\cite{trent2024bbq,veasey2024osq}, and OpenSearch exposes quantization as a storage choice~\cite{opensearch2024bq}.

\paragraph{Hashing and representation dependence}
Random-hyperplane rounding connects angular similarity to binary comparisons~\cite{goemans1995improved,charikar2002similarity}; locality-sensitive hashing~\cite{indyk1998lsh,datar2004pstable} and fast Johnson--Lindenstrauss transforms~\cite{ailon2009fastjlt} organize randomized retrieval. QuIVer instead uses coordinate sign bits directly, tying its behavior to the embedding's native geometry. InfoNCE~\cite{oord2018infonce} and the Gaussian structure it induces~\cite{betser2026infonce}, alignment and uniformity~\cite{wang2020understanding}, and the multimodal modality gap~\cite{liang2022modgap} characterize that geometry. Iterative quantization~\cite{gong2011itq} and anisotropic quantization~\cite{guo2020scann} adapt codes to search, and temperature control in contrastive training improves embedding compressibility~\cite{dinu2025temperature}; our study takes embeddings as given.

\section{Conclusion}
QuIVer combines fast graph construction, compact navigation state, and accurate retrieval through a shared two-bit representation. Integer scores rank construction candidates at low cost, while reconstruction-aware navigation preserves candidates for bounded reranking. Parallel construction and reusable bitwise kernels make this precision split effective in TriviumDB. The evaluated configurations occupy measured Pareto-frontier positions on all three core workloads, whether memory counts navigation state alone or also the original vectors. On Cohere-5M, QuIVer retains 97\% of SymphonyQG's 32-thread throughput with 54\% less index preparation time and 47\% less total logical memory, and its INT8-reranking SSD mode serves the same collection at about 5 GiB of steady-state process RSS. The 21-dataset study, held-out queries, and controlled crossovers show high recall on most text and image embeddings, place non-negative descriptors and cross-modal queries outside this range, and identify selection rules that trade construction time for graph quality.

\clearpage
\ifnum\numexpr\value{page}-1\relax>12
  \xdef\quiverLayoutIssues{\quiverLayoutIssues\space 正文页数=\number\numexpr\value{page}-1\relax\space（上限12）；}
\fi
\section*{AI-Assisted Preparation}
OpenAI Codex was used to assist with drafting and editing the manuscript text and with generating plotting and table-conversion code from the recorded experimental results. The use covers the abstract, Sections I--VI, and the associated figures and tables. OpenAI's image-generation tool supplied visual concept references for Figures~\ref{fig:pipeline} and~\ref{fig:theory-geometry}; their final illustrations retain editable vector geometry and text. The numerical-figure workflow also used the scientific-visualization skill from Scientific Agent Skills~\cite{kassis2026skills}. Experimental values were extracted from the archived measurements; the theoretical illustrations use explicitly identified conceptual examples.

\bibliographystyle{IEEEtran}
\bibliography{references}

\begin{thebibliography}{10}
\providecommand{\url}[1]{#1}
\csname url@samestyle\endcsname
\providecommand{\newblock}{\relax}
\providecommand{\bibinfo}[2]{#2}
\providecommand{\BIBentrySTDinterwordspacing}{\spaceskip=0pt\relax}
\providecommand{\BIBentryALTinterwordstretchfactor}{4}
\providecommand{\BIBentryALTinterwordspacing}{\spaceskip=\fontdimen2\font plus
\BIBentryALTinterwordstretchfactor\fontdimen3\font minus
  \fontdimen4\font\relax}
\providecommand{\BIBforeignlanguage}[2]{{%
\expandafter\ifx\csname l@#1\endcsname\relax
\typeout{** WARNING: IEEEtran.bst: No hyphenation pattern has been}%
\typeout{** loaded for the language `#1'. Using the pattern for}%
\typeout{** the default language instead.}%
\else
\language=\csname l@#1\endcsname
\fi
#2}}
\providecommand{\BIBdecl}{\relax}
\BIBdecl

\bibitem{lewis2020rag}
\BIBentryALTinterwordspacing
P.~Lewis, E.~Perez, A.~Piktus, F.~Petroni, V.~Karpukhin, N.~Goyal,
  H.~K{\"u}ttler, M.~Lewis, W.~tau Yih, T.~Rockt{\"a}schel, S.~Riedel, and
  D.~Kiela, ``Retrieval-augmented generation for knowledge-intensive {NLP}
  tasks,'' in \emph{Advances in Neural Information Processing Systems},
  vol.~33, 2020, pp. 9459--9474. [Online]. Available:
  \url{https://proceedings.neurips.cc/paper/2020/hash/6b493230205f780e1bc26945df7481e5-Abstract.html}
\BIBentrySTDinterwordspacing

\bibitem{sobot2023voyager}
\BIBentryALTinterwordspacing
P.~Sobot, ``Introducing {Voyager}: {Spotify}'s new nearest-neighbor search
  library,'' Spotify Engineering, Oct. 2023. [Online]. Available:
  \url{https://engineering.atspotify.com/2023/10/introducing-voyager-spotifys-new-nearest-neighbor-search-library}
\BIBentrySTDinterwordspacing

\bibitem{fu2025copilot}
\BIBentryALTinterwordspacing
S.~Fu, ``{GitHub Copilot} gets smarter at finding your code: Inside our new
  embedding model,'' The GitHub Blog, Sep. 2025. [Online]. Available:
  \url{https://github.blog/news-insights/product-news/copilot-new-embedding-model-vs-code/}
\BIBentrySTDinterwordspacing

\bibitem{google2026embedding}
\BIBentryALTinterwordspacing
{Google Cloud}, ``{Gemini Embedding 2},'' Model documentation, 2026, accessed:
  October 24, 2026. [Online]. Available:
  \url{https://docs.cloud.google.com/gemini-enterprise-agent-platform/models/gemini/embedding-2}
\BIBentrySTDinterwordspacing

\bibitem{segura2026structure}
J.~Segura, R.~Sanchez-Garcia, S.~Bittrich, Y.~Rose, S.~K. Burley, and J.~M.
  Duarte, ``Multi-scale structural similarity embedding search across entire
  proteomes,'' \emph{Bioinformatics}, vol.~42, no.~3, p. btag058, 2026.

\bibitem{aguerrebere2023lvq}
C.~Aguerrebere, I.~S. Bhati, M.~Hildebrand, M.~Tepper, and T.~Willke,
  ``Similarity search in the blink of an eye with compressed indices,''
  \emph{Proceedings of the VLDB Endowment}, vol.~16, no.~11, pp. 3433--3446,
  2023.

\bibitem{ngt}
\BIBentryALTinterwordspacing
{NGT Contributors}, ``{NGT}: Neighborhood graph and tree for indexing
  high-dimensional data,'' Software repository, accessed: October 23, 2026.
  [Online]. Available: \url{https://github.com/NGT-labs/NGT}
\BIBentrySTDinterwordspacing

\bibitem{gou2025symphony}
Y.~Gou, J.~Gao, Y.~Xu, and C.~Long, ``{SymphonyQG}: Towards symphonious
  integration of quantization and graph for approximate nearest neighbor
  search,'' \emph{Proceedings of the ACM on Management of Data}, vol.~3, no.~1,
  pp. 1--26, 2025.

\bibitem{trent2024bbq}
\BIBentryALTinterwordspacing
B.~Trent, ``Better binary quantization ({BBQ}) in {Lucene} and
  {Elasticsearch},'' Elastic Search Labs, Nov. 2024. [Online]. Available:
  \url{https://www.elastic.co/search-labs/blog/better-binary-quantization-lucene-elasticsearch}
\BIBentrySTDinterwordspacing

\bibitem{veasey2024osq}
\BIBentryALTinterwordspacing
T.~Veasey, ``Understanding optimized scalar quantization,'' Elastic Search
  Labs, Dec. 2024. [Online]. Available:
  \url{https://www.elastic.co/search-labs/blog/scalar-quantization-optimization}
\BIBentrySTDinterwordspacing

\bibitem{jayaram2019diskann}
\BIBentryALTinterwordspacing
S.~J. Subramanya, F.~Devvrit, H.~V. Simhadri, R.~Krishnawamy, and R.~Kadekodi,
  ``{DiskANN}: Fast accurate billion-point nearest neighbor search on a single
  node,'' in \emph{Advances in Neural Information Processing Systems}, vol.~32,
  2019. [Online]. Available:
  \url{https://proceedings.neurips.cc/paper/2019/hash/09853c7fb1d3f8ee67a61b6bf4a7f8e6-Abstract.html}
\BIBentrySTDinterwordspacing

\bibitem{wang2024starling}
M.~Wang, W.~Xu, X.~Yi, S.~Wu, Z.~Peng, X.~Ke, Y.~Gao, X.~Xu, R.~Guo, and
  C.~Xie, ``{Starling}: An {I/O}-efficient disk-resident graph index framework
  for high-dimensional vector similarity search on data segment,''
  \emph{Proceedings of the ACM on Management of Data}, vol.~2, no.~1, 2024.

\bibitem{guo2025pipeann}
\BIBentryALTinterwordspacing
H.~Guo and Y.~Lu, ``Achieving low-latency graph-based vector search via
  aligning best-first search algorithm with {SSD},'' in \emph{Proceedings of
  the 19th USENIX Symposium on Operating Systems Design and Implementation},
  2025, pp. 171--186. [Online]. Available:
  \url{https://www.usenix.org/conference/osdi25/presentation/guo}
\BIBentrySTDinterwordspacing

\bibitem{bernstein1924}
S.~N. Bernstein, ``On a modification of {Chebyshev}'s inequality and of the
  error formula of {Laplace},'' \emph{Ann. Sci. Inst. Sav. Ukraine, Sect.
  Math.}, vol.~1, pp. 38--49, 1924, in Russian.

\bibitem{boucheron2013concentration}
S.~Boucheron, G.~Lugosi, and P.~Massart, \emph{Concentration Inequalities: A
  Nonasymptotic Theory of Independence}.\hskip 1em plus 0.5em minus 0.4em\relax
  Oxford University Press, 2013.

\bibitem{vershynin2018highdim}
R.~Vershynin, \emph{High-Dimensional Probability: An Introduction with
  Applications in Data Science}.\hskip 1em plus 0.5em minus 0.4em\relax
  Cambridge University Press, 2018.

\bibitem{xiao2026decisions}
\BIBentryALTinterwordspacing
W.~Xiao and X.~Cao, ``When does low-bit quantization preserve the decisions of
  vector search?'' arXiv:2609.09854, 2026. [Online]. Available:
  \url{https://arxiv.org/abs/2609.09854}
\BIBentrySTDinterwordspacing

\bibitem{chen2024bgem3}
\BIBentryALTinterwordspacing
J.~Chen, S.~Xiao, P.~Zhang, K.~Luo, D.~Lian, and Z.~Liu, ``{M3}-embedding:
  Multi-linguality, multi-functionality, multi-granularity text embeddings
  through self-knowledge distillation,'' arXiv:2402.03216, 2024. [Online].
  Available: \url{https://arxiv.org/abs/2402.03216}
\BIBentrySTDinterwordspacing

\bibitem{woltclip2024}
\BIBentryALTinterwordspacing
{Qdrant}, ``Wolt food {CLIP ViT-B/32} embeddings,'' Dataset, accessed: October
  24, 2026. [Online]. Available:
  \url{https://huggingface.co/datasets/Qdrant/wolt-food-clip-ViT-B-32-embeddings}
\BIBentrySTDinterwordspacing

\bibitem{desai2021redcaps}
\BIBentryALTinterwordspacing
K.~Desai, G.~Kaul, Z.~Aysola, and J.~Johnson, ``{RedCaps}: Web-curated
  image-text data created by the people, for the people,'' in \emph{Proceedings
  of the Neural Information Processing Systems Track on Datasets and
  Benchmarks}, 2021. [Online]. Available:
  \url{https://arxiv.org/abs/2111.11431}
\BIBentrySTDinterwordspacing

\bibitem{aumuller2020annbenchmarks}
M.~Aum{\"u}ller, E.~Bernhardsson, and A.~Faithfull, ``{ANN-Benchmarks}: A
  benchmarking tool for approximate nearest neighbor algorithms,''
  \emph{Information Systems}, vol.~87, p. 101374, 2020.

\bibitem{aumuller2023trends}
\BIBentryALTinterwordspacing
M.~Aum{\"u}ller and M.~Ceccarello, ``Recent approaches and trends in
  approximate nearest neighbor search, with remarks on benchmarking,''
  \emph{IEEE Data Engineering Bulletin}, vol.~46, no.~3, pp. 89--105, 2023.
  [Online]. Available: \url{https://sites.computer.org/debull/A23sept/p89.pdf}
\BIBentrySTDinterwordspacing

\bibitem{li2019annanalysis}
W.~Li, Y.~Zhang, Y.~Sun, W.~Wang, M.~Li, W.~Zhang, and X.~Lin, ``Approximate
  nearest neighbor search on high dimensional data---experiments, analyses, and
  improvement,'' \emph{IEEE Transactions on Knowledge and Data Engineering},
  vol.~32, no.~8, pp. 1475--1488, 2020.

\bibitem{pennington2014glove}
\BIBentryALTinterwordspacing
J.~Pennington, R.~Socher, and C.~D. Manning, ``{GloVe}: Global vectors for word
  representation,'' in \emph{Proceedings of the 2014 Conference on Empirical
  Methods in Natural Language Processing}, 2014, pp. 1532--1543. [Online].
  Available: \url{https://aclanthology.org/D14-1162/}
\BIBentrySTDinterwordspacing

\bibitem{jaasaari2025vibe}
\BIBentryALTinterwordspacing
E.~J{\"a}{\"a}saari, V.~Hyv{\"o}nen, M.~Ceccarello, T.~Roos, and
  M.~Aum{\"u}ller, ``{VIBE}: Vector index benchmark for embeddings,''
  arXiv:2505.17810v1, 2025. [Online]. Available:
  \url{https://arxiv.org/abs/2505.17810v1}
\BIBentrySTDinterwordspacing

\bibitem{he2012difficulty}
\BIBentryALTinterwordspacing
J.~He, S.~Kumar, and S.-F. Chang, ``On the difficulty of nearest neighbor
  search,'' in \emph{Proceedings of the 29th International Conference on
  Machine Learning}, 2012, pp. 41--48. [Online]. Available:
  \url{https://arxiv.org/abs/1206.6411}
\BIBentrySTDinterwordspacing

\bibitem{aumuller2021lid}
M.~Aum{\"u}ller and M.~Ceccarello, ``The role of local dimensionality measures
  in benchmarking nearest neighbor search,'' \emph{Information Systems}, vol.
  101, p. 101807, 2021.

\bibitem{kleinberg2000smallworld}
J.~Kleinberg, ``The small-world phenomenon: An algorithmic perspective,'' in
  \emph{Proceedings of the ACM Symposium on Theory of Computing}, 2000, pp.
  163--170.

\bibitem{karger2002growth}
D.~R. Karger and M.~Ruhl, ``Finding nearest neighbors in growth-restricted
  metrics,'' in \emph{Proceedings of the ACM Symposium on Theory of Computing},
  2002, pp. 741--750.

\bibitem{malkov2020hnsw}
Y.~A. Malkov and D.~A. Yashunin, ``Efficient and robust approximate nearest
  neighbor search using hierarchical navigable small world graphs,'' \emph{IEEE
  Transactions on Pattern Analysis and Machine Intelligence}, vol.~42, no.~4,
  pp. 824--836, 2020.

\bibitem{fu2019nsg}
C.~Fu, C.~Xiang, C.~Wang, and D.~Cai, ``Fast approximate nearest neighbor
  search with the navigating spreading-out graph,'' \emph{Proceedings of the
  VLDB Endowment}, vol.~12, no.~5, pp. 461--474, 2019.

\bibitem{zhu2021monotonic}
\BIBentryALTinterwordspacing
D.~Zhu and M.~Zhang, ``Understanding and generalizing monotonic proximity
  graphs for approximate nearest neighbor search,'' arXiv:2107.13052, 2021.
  [Online]. Available: \url{https://arxiv.org/abs/2107.13052}
\BIBentrySTDinterwordspacing

\bibitem{xiang2025emg}
\BIBentryALTinterwordspacing
L.~Xiang, J.~Feng, Z.~Yin, Z.~Li, D.~Xue, H.~Qin, R.~Li, and G.~Wang,
  ``$\delta$-{EMG}: A monotonic graph index for approximate nearest neighbor
  search,'' arXiv:2511.16921, 2025. [Online]. Available:
  \url{https://arxiv.org/abs/2511.16921}
\BIBentrySTDinterwordspacing

\bibitem{faissbinary}
\BIBentryALTinterwordspacing
{Faiss Contributors}, ``Binary indexes,'' Faiss documentation, accessed:
  October 23, 2026. [Online]. Available:
  \url{https://github.com/facebookresearch/faiss/wiki/Binary-indexes}
\BIBentrySTDinterwordspacing

\bibitem{wang2021graphsurvey}
M.~Wang, X.~Xu, Q.~Yue, and Y.~Wang, ``A comprehensive survey and experimental
  comparison of graph-based approximate nearest neighbor search,''
  \emph{Proceedings of the VLDB Endowment}, vol.~14, no.~11, pp. 1964--1978,
  2021.

\bibitem{wang2025flash}
M.~Wang, H.~Wu, X.~Ke, Y.~Gao, Y.~Zhu, and W.~Zhou, ``Accelerating graph
  indexing for {ANNS} on modern {CPUs},'' \emph{Proceedings of the ACM on
  Management of Data}, vol.~3, no.~3, pp. 1--29, 2025.

\bibitem{dong2026pipnn}
\BIBentryALTinterwordspacing
T.~Rubel, R.~Wen, L.~Dhulipala, L.~Gottesb{\"u}ren, R.~Jayaram, and
  J.~{\L}\k{a}cki, ``{PiPNN}: Ultra-scalable graph-based nearest neighbor
  indexing,'' arXiv:2602.21247, 2026. [Online]. Available:
  \url{https://arxiv.org/abs/2602.21247}
\BIBentrySTDinterwordspacing

\bibitem{gong2025shg}
Z.~Gong, Y.~Zeng, and L.~Chen, ``Accelerating approximate nearest neighbor
  search in hierarchical graphs: Efficient level navigation with shortcuts,''
  \emph{Proceedings of the VLDB Endowment}, vol.~18, no.~10, pp. 3518--3530,
  2025.

\bibitem{lu2026pag}
\BIBentryALTinterwordspacing
K.~Lu, Z.~Pan, J.~Qin, Y.~Ishikawa, and C.~Xiao, ``Approximate nearest neighbor
  search for modern {AI}: A projection-augmented graph approach,''
  arXiv:2603.06660, 2026. [Online]. Available:
  \url{https://arxiv.org/abs/2603.06660}
\BIBentrySTDinterwordspacing

\bibitem{zhong2025vsag}
X.~Zhong, H.~Li, J.~Jin, M.~Yang, D.~Chu, X.~Wang, Z.~Shen, W.~Jia, G.~Gu,
  Y.~Xie, X.~Lin, H.~T. Shen, J.~Song, and P.~Cheng, ``{VSAG}: An optimized
  search framework for graph-based approximate nearest neighbor search,''
  \emph{Proceedings of the VLDB Endowment}, vol.~18, no.~12, pp. 5017--5030,
  2025.

\bibitem{usearch2024}
\BIBentryALTinterwordspacing
A.~Vardanian, ``{USearch}: Smaller \& faster single-file similarity search
  engine for vectors \& strings,'' Software repository, accessed: October 24,
  2026. [Online]. Available: \url{https://github.com/unum-cloud/usearch}
\BIBentrySTDinterwordspacing

\bibitem{yue2024rpq}
\BIBentryALTinterwordspacing
Q.~Yue, X.~Xu, Y.~Wang, Y.~Tao, and X.~Luo, ``Routing-guided learned product
  quantization for graph-based approximate nearest neighbor search,'' in
  \emph{Proceedings of the IEEE International Conference on Data Engineering},
  2024, pp. 4870--4883. [Online]. Available:
  \url{https://arxiv.org/abs/2311.18724}
\BIBentrySTDinterwordspacing

\bibitem{jegou2011pq}
H.~J{\'e}gou, M.~Douze, and C.~Schmid, ``Product quantization for nearest
  neighbor search,'' \emph{IEEE Transactions on Pattern Analysis and Machine
  Intelligence}, vol.~33, no.~1, pp. 117--128, 2011.

\bibitem{ge2013opq}
T.~Ge, K.~He, Q.~Ke, and J.~Sun, ``Optimized product quantization for
  approximate nearest neighbor search,'' in \emph{Proceedings of the IEEE
  Conference on Computer Vision and Pattern Recognition}, 2013, pp. 2946--2953.

\bibitem{johnson2021faiss}
J.~Johnson, M.~Douze, and H.~J{\'e}gou, ``Billion-scale similarity search with
  {GPUs},'' \emph{IEEE Transactions on Big Data}, vol.~7, no.~3, pp. 535--547,
  2021.

\bibitem{matsui2018pqs}
Y.~Matsui, Y.~Uchida, H.~J{\'e}gou, and S.~Satoh, ``A survey of product
  quantization,'' \emph{ITE Transactions on Media Technology and Applications},
  vol.~6, no.~1, pp. 2--10, 2018.

\bibitem{andre2015pqfs}
F.~Andr{\'e}, A.-M. Kermarrec, and N.~{Le Scouarnec}, ``Cache locality is not
  enough: High-performance nearest neighbor search with product quantization
  fast scan,'' \emph{Proceedings of the VLDB Endowment}, vol.~9, no.~4, pp.
  288--299, 2015.

\bibitem{andre2017quickadc}
------, ``Accelerated nearest neighbor search with quick {ADC},'' in
  \emph{Proceedings of the ACM International Conference on Multimedia
  Retrieval}, 2017, pp. 159--166.

\bibitem{gao2024rabitq}
J.~Gao and C.~Long, ``{RaBitQ}: Quantizing high-dimensional vectors with a
  theoretical error bound for approximate nearest neighbor search,''
  \emph{Proceedings of the ACM on Management of Data}, vol.~2, no.~3, pp.
  167:1--167:27, 2024.

\bibitem{gao2025extended}
J.~Gao, Y.~Gou, Y.~Xu, Y.~Yang, C.~Long, and R.~C.-W. Wong, ``Practical and
  asymptotically optimal quantization of high-dimensional vectors in euclidean
  space for approximate nearest neighbor search,'' \emph{Proceedings of the ACM
  on Management of Data}, vol.~3, no.~3, pp. 1--26, 2025.

\bibitem{opensearch2024bq}
\BIBentryALTinterwordspacing
{OpenSearch Project}, ``Vector quantization,'' OpenSearch documentation,
  accessed: October 24, 2026. [Online]. Available:
  \url{https://docs.opensearch.org/latest/vector-search/optimizing-storage/knn-vector-quantization/}
\BIBentrySTDinterwordspacing

\bibitem{goemans1995improved}
M.~X. Goemans and D.~P. Williamson, ``Improved approximation algorithms for
  maximum cut and satisfiability problems using semidefinite programming,''
  \emph{Journal of the ACM}, vol.~42, no.~6, pp. 1115--1145, 1995.

\bibitem{charikar2002similarity}
M.~S. Charikar, ``Similarity estimation techniques from rounding algorithms,''
  in \emph{Proceedings of the ACM Symposium on Theory of Computing}, 2002, pp.
  380--388.

\bibitem{indyk1998lsh}
P.~Indyk and R.~Motwani, ``Approximate nearest neighbors: Towards removing the
  curse of dimensionality,'' in \emph{Proceedings of the ACM Symposium on
  Theory of Computing}, 1998, pp. 604--613.

\bibitem{datar2004pstable}
M.~Datar, N.~Immorlica, P.~Indyk, and V.~S. Mirrokni, ``Locality-sensitive
  hashing scheme based on p-stable distributions,'' in \emph{Proceedings of the
  ACM Symposium on Computational Geometry}, 2004, pp. 253--262.

\bibitem{ailon2009fastjlt}
N.~Ailon and B.~Chazelle, ``The fast {Johnson--Lindenstrauss} transform and
  approximate nearest neighbors,'' \emph{SIAM Journal on Computing}, vol.~39,
  no.~1, pp. 302--322, 2009.

\bibitem{oord2018infonce}
\BIBentryALTinterwordspacing
A.~van~den Oord, Y.~Li, and O.~Vinyals, ``Representation learning with
  contrastive predictive coding,'' arXiv:1807.03748, 2018. [Online]. Available:
  \url{https://arxiv.org/abs/1807.03748}
\BIBentrySTDinterwordspacing

\bibitem{betser2026infonce}
\BIBentryALTinterwordspacing
R.~Betser, E.~Gofer, M.~Y. Levi, and G.~Gilboa, ``{InfoNCE} induces {Gaussian}
  distribution,'' in \emph{Proceedings of the International Conference on
  Learning Representations}, 2026. [Online]. Available:
  \url{https://openreview.net/forum?id=BlSH7gNQSq}
\BIBentrySTDinterwordspacing

\bibitem{wang2020understanding}
\BIBentryALTinterwordspacing
T.~Wang and P.~Isola, ``Understanding contrastive representation learning
  through alignment and uniformity on the hypersphere,'' in \emph{Proceedings
  of the International Conference on Machine Learning}, ser. Proceedings of
  Machine Learning Research, vol. 119, 2020, pp. 9929--9939. [Online].
  Available: \url{https://proceedings.mlr.press/v119/wang20k.html}
\BIBentrySTDinterwordspacing

\bibitem{liang2022modgap}
\BIBentryALTinterwordspacing
V.~W. Liang, Y.~Zhang, Y.~Kwon, S.~Yeung, and J.~Y. Zou, ``Mind the gap:
  Understanding the modality gap in multi-modal contrastive representation
  learning,'' in \emph{Advances in Neural Information Processing Systems},
  vol.~35, 2022, pp. 17\,612--17\,625. [Online]. Available:
  \url{https://proceedings.neurips.cc/paper_files/paper/2022/hash/702f4db7543a7432431df588d57bc7c9-Abstract-Conference.html}
\BIBentrySTDinterwordspacing

\bibitem{gong2011itq}
Y.~Gong and S.~Lazebnik, ``Iterative quantization: A procrustean approach to
  learning binary codes,'' in \emph{Proceedings of the IEEE Conference on
  Computer Vision and Pattern Recognition}, 2011, pp. 817--824.

\bibitem{guo2020scann}
\BIBentryALTinterwordspacing
R.~Guo, P.~Sun, E.~Lindgren, Q.~Geng, D.~Simcha, F.~Chern, and S.~Kumar,
  ``Accelerating large-scale inference with anisotropic vector quantization,''
  in \emph{Proceedings of the International Conference on Machine Learning},
  ser. Proceedings of Machine Learning Research, vol. 119, 2020, pp.
  3887--3896. [Online]. Available:
  \url{https://proceedings.mlr.press/v119/guo20h.html}
\BIBentrySTDinterwordspacing

\bibitem{dinu2025temperature}
\BIBentryALTinterwordspacing
G.~Dinu, C.~Barrett, Y.~Xiang, M.~R. Calvo, A.~Currey, and X.~Niu, ``Effective
  post-training embedding compression via temperature control in contrastive
  training,'' in \emph{Proceedings of the International Conference on Learning
  Representations}, 2025. [Online]. Available:
  \url{https://proceedings.iclr.cc/paper_files/paper/2025/hash/1d5a92867cf463fad136cfa23395840b-Abstract-Conference.html}
\BIBentrySTDinterwordspacing

\bibitem{kassis2026skills}
\BIBentryALTinterwordspacing
T.~Kassis, V.~Agarwal, Y.~He, D.~Patel, and A.~M. Brueckner, ``Scientific agent
  skills: A library of procedural knowledge for research agents,'' 2026.
  [Online]. Available: \url{https://arxiv.org/abs/2609.00065}
\BIBentrySTDinterwordspacing

\end{thebibliography}
\end{document}